\documentclass[journal,twoside,web]{ieeecolor}
\usepackage{generic}
\usepackage{cite}
\usepackage{amsmath,amssymb,amsfonts}
\usepackage{algorithmic}
\usepackage{graphicx}
\usepackage{textcomp}
\usepackage{xspace}
\usepackage{hyperref}
\usepackage{array}
\usepackage{tabularx}
\usepackage{siunitx}
\usepackage{booktabs}
\usepackage{svg}
\usepackage{amsmath}
\usepackage{soul}
\usepackage{ulem}

\usepackage[T1]{fontenc}
\def\t{\ensuremath{\mathcal{T}}\xspace} 
\def\r{\ensuremath{\mathcal{R}}\xspace} 
\def\ctpat{$\text{CT}_{pat}$\xspace}

\newcommand{\revisionAdd}[1]{\textcolor{black}{#1}}
\newcommand{\revisionDel}[1]{\unskip}

\def\BibTeX{{\rm B\kern-.05em{\sc i\kern-.025em b}\kern-.08em
    T\kern-.1667em\lower.7ex\hbox{E}\kern-.125emX}}
\begin{document}

\makeatletter

\newcommand{\thickhline}{%
    \noalign {\ifnum 0=`}\fi \hrule height 1pt
    \futurelet \reserved@a \@xhline
}
\newcolumntype{"}{@{\hskip\tabcolsep\vrule width 1pt\hskip\tabcolsep}}
\makeatother
\title{Pose-dependent weights and Domain Randomization for fully automatic X-ray to CT Registration}



\author{Matthias Grimm* , Javier Esteban, Mathias Unberath and Nassir Navab
\thanks{M. Grimm and J. Esteban contributed equally to this work. Asterisk indicates corresponding author. This work was supported by the German Federal Ministry of Research and Education (FKZ: 13GW0236B) }
\thanks{M. Grimm, J. Esteban and N. Navab are with Computer Aided Medical Procedures (CAMP), Technische Universit\"at M\"unchen, Germany (email: matthias.grimm|javier.esteban|nassir.navab@tum.de}
\thanks{M. Unberath is with Laboratory for Computational Sensing + Robotics, Johns Hopkins University, Baltimore, MD, United States (e-mail: \mbox{unberath@jhu.edu)}}
\thanks{Preprint, to appear in IEEE Transactions on Medical Imaging (https://doi.org/10.1109/TMI.2021.3073815).
}
}

\maketitle
\begin{abstract}
Fully automatic X-ray to CT registration requires a solid initialization to provide an initial alignment within the capture range of existing intensity-based registrations.
This work adresses that need by providing a novel automatic initialization, which enables end to end registration.
First, a neural network is trained once to detect a set of anatomical landmarks on simulated X-rays.
A domain randomization scheme is proposed to enable the network to overcome the challenge of being trained purely on simulated data and run inference on real X-rays.
Then, for each patient CT, a fully-automatic patient-specific landmark extraction scheme is used.
It is based on backprojecting and clustering the previously trained network's predictions on a set of simulated X-rays.
Next, the network is retrained to detect the new landmarks.
Finally the combination of network and 3D landmark locations is used to compute the initialization using a perspective-n-point algorithm.
During the computation of the pose, a weighting scheme is introduced to incorporate the confidence of the network in detecting the landmarks.
The algorithm is evaluated on the pelvis using both real and simulated x-rays. The mean ($\pm$ standard deviation) target registration error in millimetres is $4.1\pm4.3$ for simulated X-rays with a success rate of $92\%$ and $4.2\pm3.9$ for real X-rays with a success rate of $86.8\%$, where a success is defined as a translation error of less than $30~mm$.
\end{abstract}


\section{Introduction}
\label{sec:introduction}
\IEEEPARstart{I}{n} image-guided therapy, preoperative medical scans such as computed tomography (CT) or magnetic resonance imaging are used to plan and guide a surgery.
However, the planned information is relative to the coordinate system of the preoperative data.
In order to utilize it during the intervention,  a mapping~\t, called a registration,  between the coordinate frame of the preoperative data and some intraoperative modality, such as X-ray or Ultrasound, is computed.
Automatic registration is crucial for enabling image-guided therapy and therefore reducing the mental load and improving the clinical outcome.
\par
An important property of a registration algorithm is its capture range, namely the maximum initial displacement for which it is still able to perform a successful registration~\cite{markelj2012review}.
In other words, for a given registration method~\r the initial offset between two given scans $\mathbf{I}_1$ and $\mathbf{I}_2$ must be inside the capture range, otherwise the alignment operation $\r(\mathbf{I}_1, \mathbf{I}_2)$ is likely to fail.
Hence a fully-automatic registration method requires two components: An initialization method~$\mathcal{I}$ whose goal is to align the two input scans such that they fall into the capture range of~\r and a subsequent fine-grain registration method~\r whose goal is to perform the final alignment.
So far most previous research has focused on~\r, while $\mathcal{I}$ still remains the Achilles Heel of most registration methods. Marker-based approaches do not require an initialization, however, they suffer from other problems~\cite{markelj2012review}. Bone-implanted markers require a second, invasive procedure, whereas skin-attached markers are susceptible to deformations and hence result in lower accuracy~\cite{markelj2012review}. Purely image-based methods, such as intensity-based or feature-based, require a correct initialization.\par
\begin{figure*}
  \centering

\includegraphics[width=0.9\textwidth]{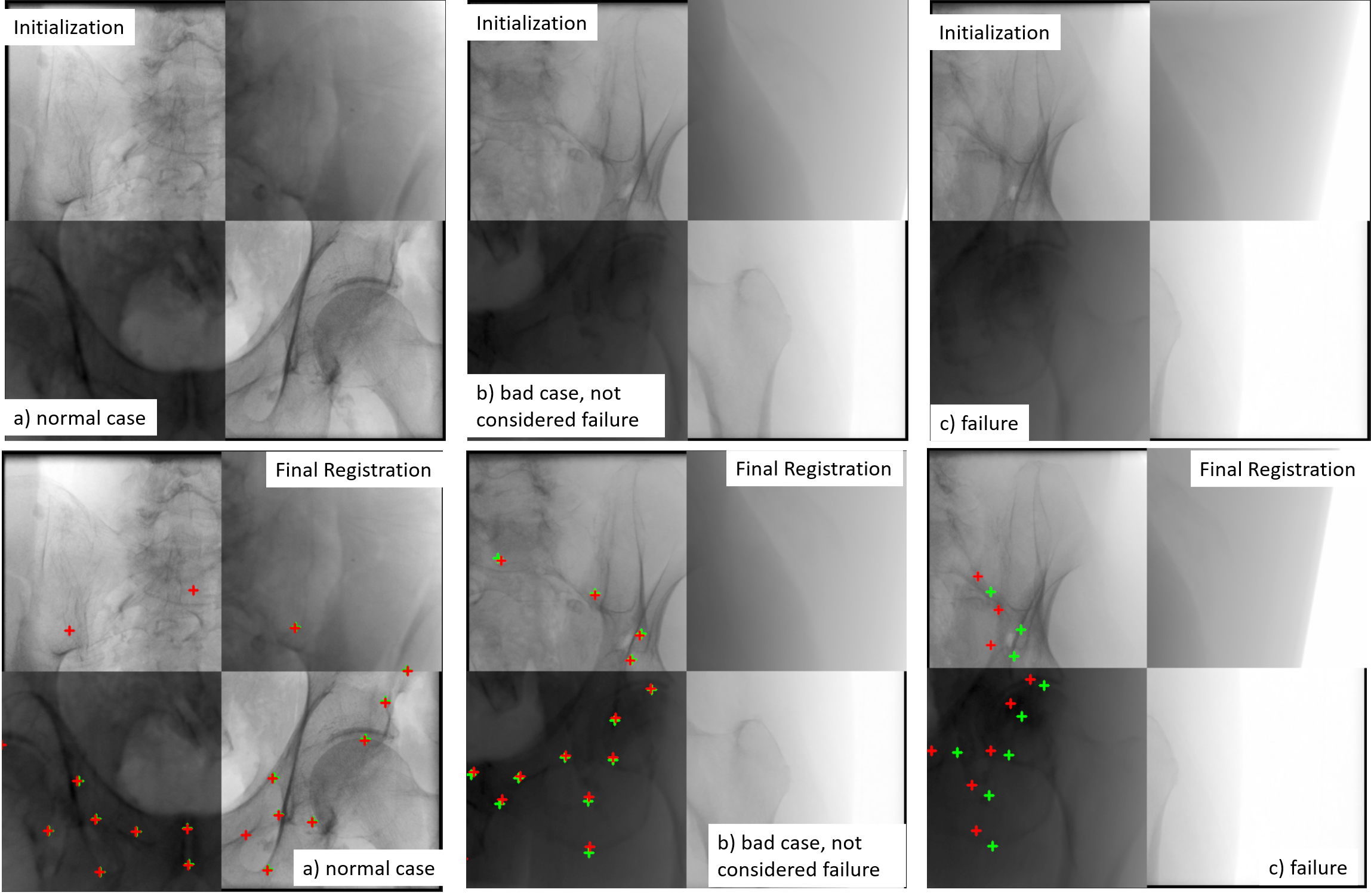}
\caption{
Qualitative results for the initialization and final registration for $\textbf{CT}_{cad} - 2$. a) Representative case. b) bad case, not considered failure. c) Failure case (the computed pose is on the wrong side of the patient). Most failures fall into that category.
}
\label{fig_chessboard}
\end{figure*}
Although registration methods have been widely researched, providing a robust initialization remains an open problem. A method using the projection-slice theorem and phase correlation was proposed~\cite{van2010robust}. Despite having good results, the authors mention two drawbacks: the runtime was not suited for intraoperative use, and there were geometric discrepancies between the theoretical parallel-beam and the actual cone-beam. A regression technique was designed to handle arbitrary motion and predict slice transformations relative to a learned canonical atlas coordinate frame~\cite{hou2017predicting}. The method is however limited by its relatively poor accuracy. Recently, a reinforcement learning approach~\cite{miao2018dilated} using a convolutional neural network as a policy to perform registration in a Markov Decision Process was proposed. However, the method requires an initialization and has a restricted capture range.
\cite{miao2016real} propose\revisionAdd{s} a library based method to recover the pose of an articulated object in X-ray images. The method relies on decoupling the registration process into estimating first parameters affecting the geometry and then parameters not affecting the geometry of the object in question. However, the method was only evaluated on transesophageal echocardiogram probes, not on actual human anatomies.
Another method is based on a combination of multiple methods to register the individual anatomical parts of the Pelvis to a preoperative CT~\cite{grupp2019automatic}. While the intraoperative part of the method is automatic, it still requires manual annotation of the CT.
\begin{figure*}
\includegraphics[width=1\textwidth]{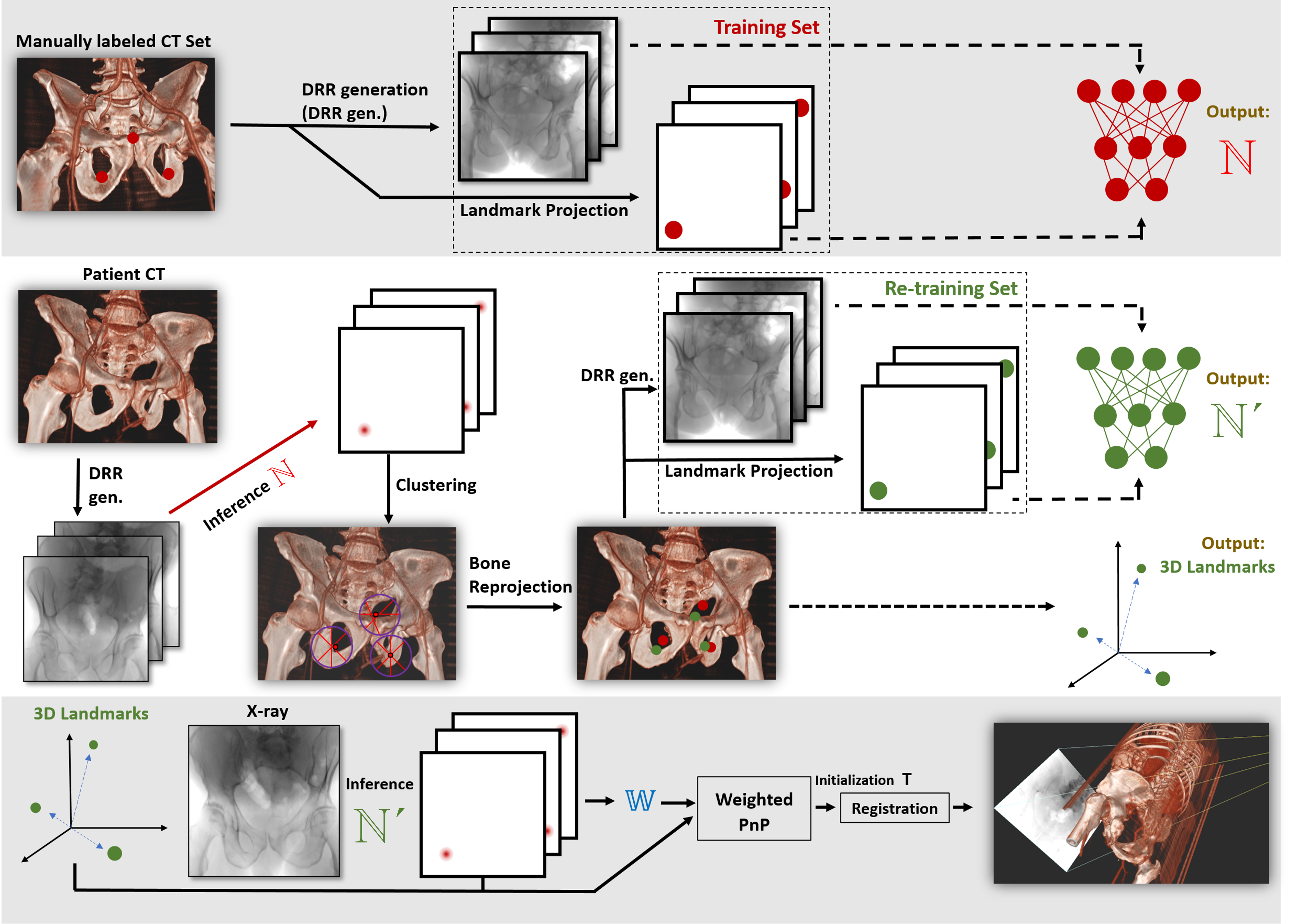}
\caption{
Schematic overview over the framework.
For visual purposes, only three of the 23 landmarks are depicted.
}
\label{fig_workflow}
\end{figure*}
\begin{figure}
  \centering
\includegraphics[width=0.4\textwidth]{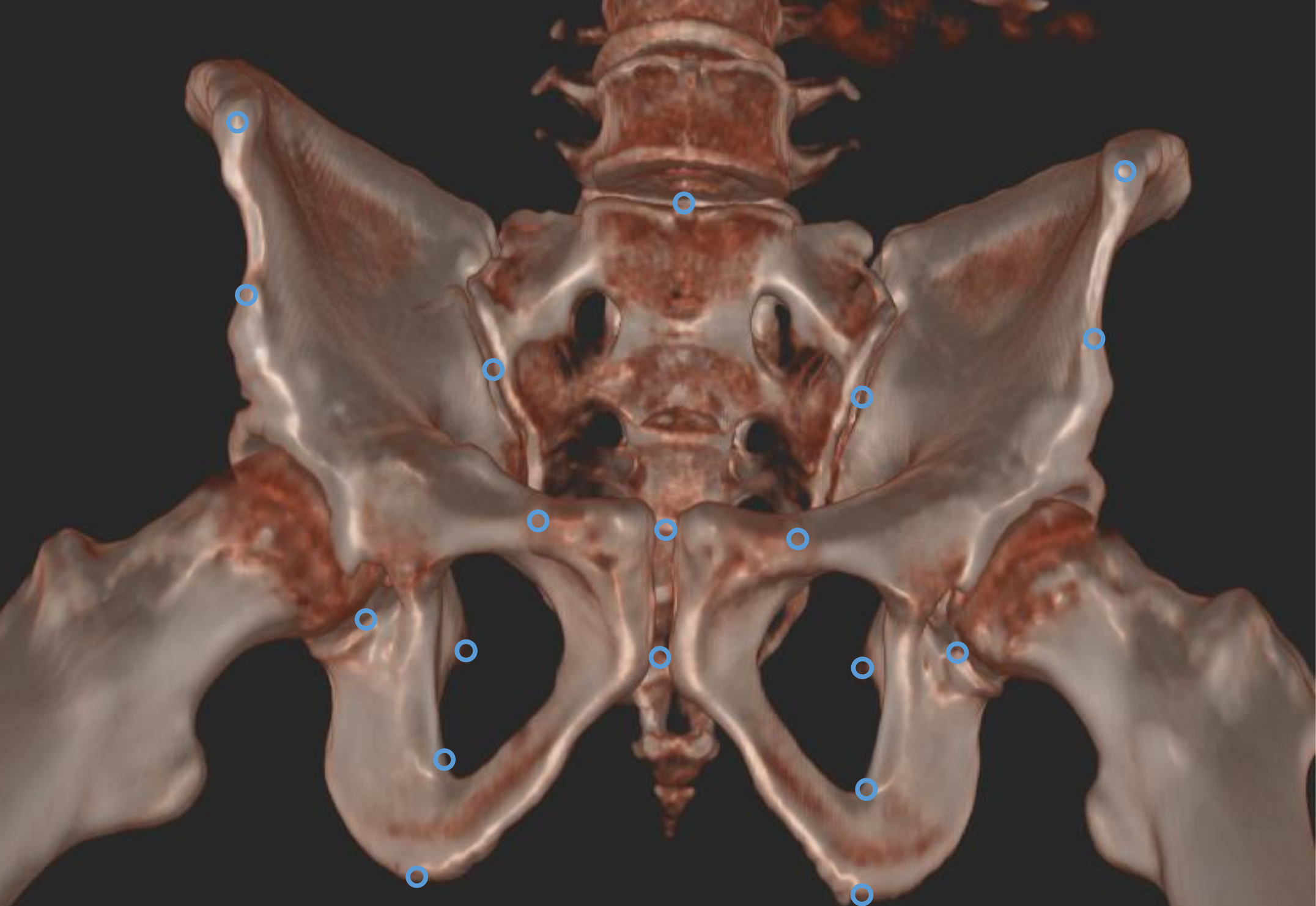}
\caption{19 of the 23 3D landmarks $m_i$, whose projections are used for training the network $\mathbb{N}$. The landmarks are the following points: Left and right inferior ischial tuberosity, left and right inferior obturator foramen, left and right pubic tubercle, left and right superior obturator foramen, left and right acetabular notch, left and right center of femoral head, left and right ischial spine, left and right sacroiliac joint, left and right anterior inferior iliac spine, left and right anterior superior iliac spine, \revisionDel{the} center of \revisionAdd{the} sacral promontory and \revisionDel{the} inferior and superior pubis symphisis.
}\label{fig_CT}
\end{figure}
A different strategy relying on an initialization framework using multiple coarse segmentations was proposed~\cite{rackerseder2018initialize}.
A drawback of this method is that the initialization step is requiring a third-party segmentation algorithm that has to be provided by the user.
An original approach was recently introduced to automatically detect anatomical landmarks in 2D~X-ray images. This allows to determine the transformation~\t with respect to the corresponding 3D~CT volume~\cite{bier2018xray}.
Nevertheless, this method has two limitations:
1)~the original 3D~CT landmarks must be manually annotated for each new patient, which is a tedious and analyst-dependent task, and
2)~the landmark model is pre-generated from a set of patients and may therefore not fully capture patient-specific anatomical details when applied intraoperatively.
The aim of the present work is to propose a method to estimate the initial rigid transformation~\t between a preoperative 3D CT volume and intraoperative 2D X-ray images.\\
\indent Another important topic this paper addresses is bridging the domain gap between simulated and real X-rays and therefore enable neural networks purely trained on simulated X-rays, to perform successful inference on real X-rays. One main family of works deals with simulating the image formation process of X-rays as realistic as possible~\cite{unberath2018deepdrr, unberath2019enabling}. While these works show great promise, they have two main drawbacks: First, scanners usually apply a set of post-processing steps to the acquired scans. In order to simulate images as good as possible these steps need to be incorporated in the simulation process. However, the list of steps is usually unknown and differs for each scanner. Second, \revisionDel{the} experimental results suggest that networks trained purely on simulated data perform four times worse when running on real data, as opposed to running on simulated data.
Another important branch of works are based on generative adversarial networks (GANs). \cite{dou2019pnp} propose\revisionAdd{s} a GAN-based scheme for adapting segmentation networks between different modalities of medical images.  \cite{zhang2020unsupervised} use\revisionAdd{s} a task driven GAN to enable a network trained on synthetic X-rays to parse real X-rays. Unlike the proposed approach, these methods require the presence of a large collection of real X-rays in order to be trained. Since these are data-driven methods, their performance depends on the collection of real X-rays being suitable for the task at hand.\\
\indent This paper is a journal extension of a previous conference submission~\cite{esteban2019XrayReg}. The main contribution of the previous submission is a patient-specific landmark-refinement scheme based on deep learning and projective geometry. Thereby, the previous work adresses the limitations mentioned above and propose a framework, which is fully-automatic, falls into the capture range of standard registration methods and has a runtime compatible with intraoperative applications.\\
\indent This paper extends the previous work in three ways. First, a cadaver study was conducted to evaluate the algorithm on real X-rays. Second, a novel pose-dependent weighting scheme to improve the Perspective-n-Point (PnP) solver using the confidence of the trained network in detecting the landmarks is introduced and third a new augmentation scheme is used to improve the neural network's transfer from simulated to real data. Experiments are carried out on pelvis anatomy in the context of trauma surgery.

\section{Methods}
\label{sec:methods}
An overview over the proposed method is shown in Figure~\ref{fig_workflow}. The method consists of three phases, which are described hereafter.\par
\textbf{Phase \texttt{1} --- Network pre-training:}
This phase is performed only once, before the algorithm is rolled out.
Here, a convolutional neural network~$\mathbb{N}$ is trained to detect the projections of a set of~$\Omega=23$ 3D anatomical landmarks $\mathcal{M} = [m_1, m_2, \dots, m_\Omega]$ onto X-ray images using a previously published method~\cite{bier2018xray}. The 3D landmark locations, shown in Fig.~\ref{fig_CT}, were chosen such that they correspond to clinically meaningful and clearly identifiable points~\cite{bier2018xray}.

\begin{figure*}[h]
\includegraphics[width=1\textwidth]{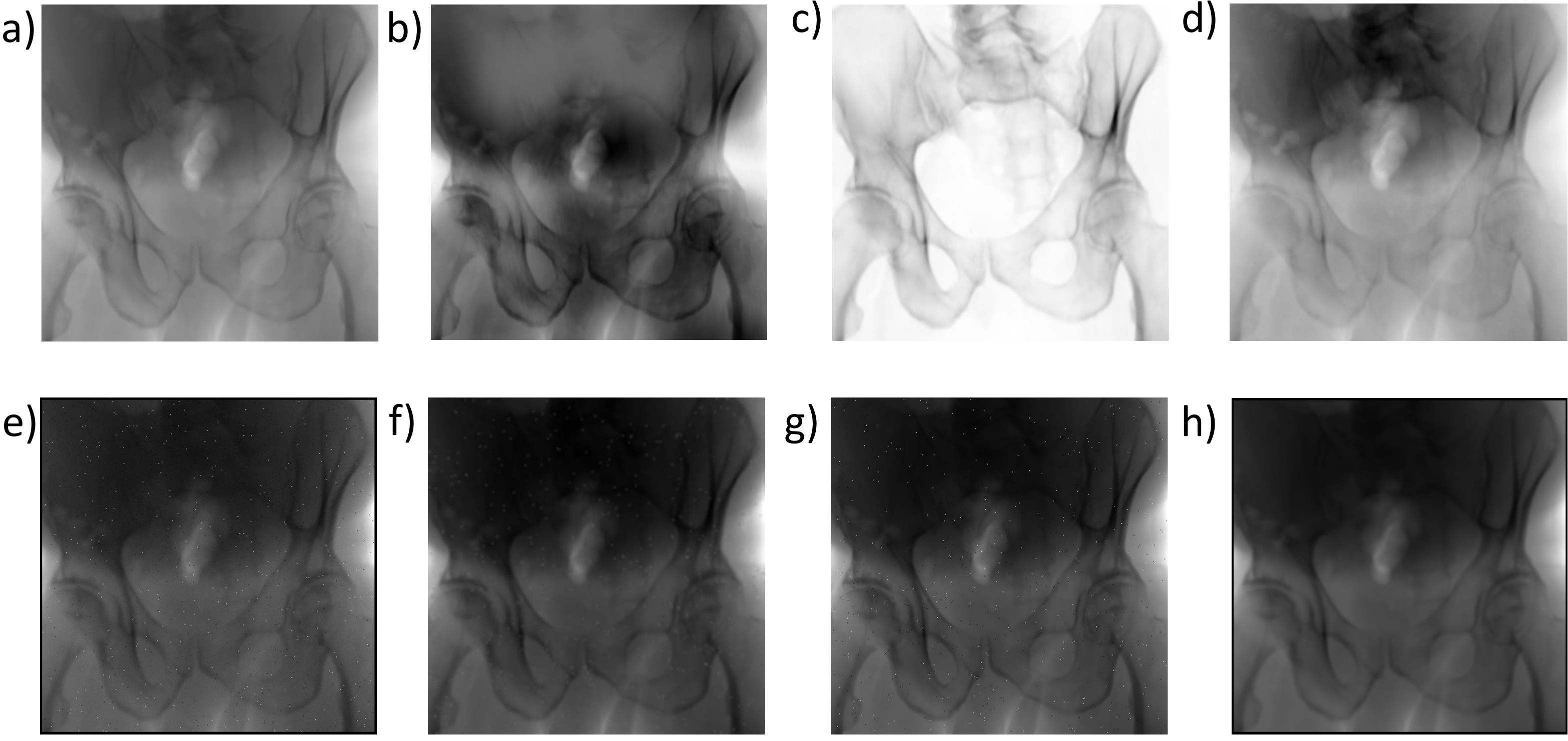}
\caption{a)-d): Images of the same pose generated using the four DRR generators used. a): DeepDRR without scatter estimation, b): DeepDRR with scatter estimation, c): publicly available raycaster ,d): commercial DRR generator. e-h): Four examples of the augmentation algorithm applied to a).}
\label{fig_drrs}
\end{figure*}

Since X-rays and CTs share the same modality, it is possible to generate reasonably realistic simulated X-rays (i.e. digitally reconstructed radiographs; DRRs) from CT scans. This allows to train neural networks purely on simulated DRRs and then enable them to bridge the domain gap to real X-rays. Thereby, the burden of obtaining large amounts of manually labelled data can be alleviated, by labelling few CTs and then projecting the labels onto a large amount of simulated X-rays. However, bridging the domain gap is not a trivial task. A recent study~\cite{unberath2019enabling} shows that even when using state of the art DRR generators, the error for anatomical
landmark detection increases by a factor of four when inference is done on real X-rays, as opposed to DRRs for a network purely trained on DRRs.\\
\indent \textit{Domain Randomization:} Building upon the work of~\cite{unberath2019enabling}, this method proposes a new technique to bridge the domain gap, inspired by the domain randomization~\cite{tobin2017domain} works in robotics. During training, the poses in the training set are generated using multiple different DRR generators $\mathbf{Gens} = \{ \mathbf{Gen_1}, \mathbf{Gen_2},..., \mathbf{Gen_n} \}$, which are not necessarily state of the art.  The intuition behind this is that each of the generators constitutes a visual style and the network will be exposed to an infinite amount of styles during training, due to the randomization. Thereby, the network will be forced to learn to deal with a large variance of styles, where the difference between the styles is larger than the domain gap between real and virtual X-ray scans. To the network, the real X-rays will appear as yet another one of these styles. This can also allow a network to become more robust to other variations, for example air cavities due to pneumoperitoneum, artifacts, or small metallic objects.
Four DRR generators, depicted in Fig.~\ref{fig_drrs}, are used. The first generator (Fig.~\ref{fig_drrs} a)) $\mathbf{Gen_1}$, corresponds to the method~\cite{unberath2018deepdrr} without scatter estimation. The second generator (Fig.~\ref{fig_drrs} b)) $\mathbf{Gen_2}$, corresponds to the same method with scatter estimation. It should be noted the publicly available implementation\footnote{\url{https://github.com/mathiasunberath/DeepDRR}} was used which does not include the final log conversion (i.e. intensities represent energy arriving at the detector).
The third DRR generator (Fig.~\ref{fig_drrs} c)) $\mathbf{Gen_3}$, corresponds to a publicly available ray caster\footnote{\url{https://github.com/SeverineHabert/DRR-renderer}}. The fourth DRR generator (Fig.~\ref{fig_drrs} d)) $\mathbf{Gen_4}$, corresponds to a commercially available DRR generator\footnote{ImFusion GmbH, Munich, Germany (\url{https://www.imfusion.de)}}. After each DRR is generated, an elaborate randomized post-processing scheme is introduced to create a further variety of input scans.\par
\textit{Post-processing scheme:} During training, before each forward pass, it is decided with probability $50\%$ whether the post-processing is done or not. If so, there are nine stages. The order of the stages is random, and each one is conducted with probability $50\%$. Each stage takes the output of the previous stage as input. After each stage, intensity values below zero are mapped to zero. The stages are as follows:
\color{black} 

\begin{itemize}
 \item \textbf{Smoothing:} $3\times3$ or $5\times5$ kernel with 50$\%$ alternation
 \item \textbf{Offset} offset sampled uniformly in the range $(-0.2 \cdot max, 0.15 \cdot max)$
 \item \textbf{Linear Scaling:} scaling factor is sampled uniformly in the range $(0.8, 1.15)$.
 \item \textbf{Renormalization:} Lower and upper bound are sampled from the intervals: $[-0.04\cdot max, 0.02\cdot max]$ and $[0.9\cdot max, 1.05\cdot max]$ 
 \item \textbf{Non-linar pixel-wise offset:}  $offset = a \cdot sin(b\cdot x + c)$. $a$ and $b$ are sampled uniformly from the interval $(0.9, 1.05)$ and $c$ is sampled uniformly from the interval $(-0.4, 0.4)$
 \item \textbf{Salt and pepper noise:} number of perturbed pixels is uniformly sampled from:  $(0.02\cdot numPixels, 0.04 \cdot numPixels)$, where $numPixels$ is the number of pixels
 \item \textbf{Gaussian noise:} Mean is sampled from $(-0.15\cdot max, 0.1\cdot max)$
 \item \textbf{Poisson noise}
 \item \textbf{Non linear Scaling:} $a \cdot sin(b \cdot x + c)$ a and b are sampled uniformly from  $(0.8,1.1)$ and c from ($-0.5, 0.5)$
\end{itemize}

where $x$ is the intensity value of the pixel and $max$ is the maximum intensity present in the image.

It can be noted that the mean of most intervals is not zero. This is to account for the fact that the real X-rays appear darker than DRRs (as can be seen in Fig.~\ref{fig_impainting}).\par
\textit{Training scheme:} Ground truth landmark locations for the training were obtained by projecting the points $m_\omega$ from CT to X-ray, using the known extrinsics and intrinsics of the virtual C-arm.
The network~$\mathbb{N}$ corresponds to a previously introduced architecture called the convolutional pose machine (CPM)~\cite{wei2016convolutional}. This architecture receives an image as input and outputs a set of heatmaps, one per landmark. A key feature of the architecture is that it is a concatenation of several convolutional neural networks, where each operates on the output of the previous one, thereby refining its estimates. For this work, only the output of the last stage is considered.
The ground truth consists of a heatmap per landmark, with a Gaussian placed at the location of the landmark. A landmark is assumed to be detected, if the maximum response of the corresponding heatmap in the output of the network is above a confidence threshold $\mu$. Input scans are normalized to the intensity interval $[0,1]$. The input image size is $512$ by $512$ pixels. Outputs are upsampled to size $512$ by $512$ using bilinear interpolation. A nice property of the CPM when compared to other architectures, such as the stacked hourglass~\cite{newell2016stacked}, is its relatively simple convergence behaviour, making it easy to train the network automatically. In general, any other network for anatomical landmark prediction could have been used for this work as long as it has a sufficiently good accuracy.\par
The training dataset, denoted as $DRR^{CT-train}_{train}$, is using a set of poses called $S^1$ to generate DRRs from a collection of CTs denoted as $\textbf{CT}_{train}$. Similarly, the validation set ($DRR^{CT-val}_{val}$) is generated by applying the same poses to a different set of CTs ($\textbf{CT}_{val}$).\par
Training was carried out with the Adam optimizer for $4$ epochs until convergence was reached, with a learning rate of $0.00001$ and a batch size of $1$, similar to~\cite{bier2018xray}.\par
\textbf{Phase \texttt{2} --- Automatic patient-specific landmarks extraction:}
This phase's inputs are a patient-specific CT, referred to as \ctpat, and the neural network $\mathbb{N}$ trained in the previous step. During this phase, a new set of anatomical 3D landmarks $\mathcal{M}' = [m_{1}', m_{2}', \dots, m_{\Omega}']$ on \ctpat is \revisionAdd{automatically} computed \revisionAdd{since the location of $\mathcal{M}$ on \ctpat is not known}. Furthermore, $\mathbb{N}$ is being retrained, yielding a new network $\mathbb{N}'$, which is able to detect the projections of $\mathcal{M}'$ on X-rays from \ctpat.
\begin{figure}[b]
\centering
\includegraphics[width=0.4\textwidth]{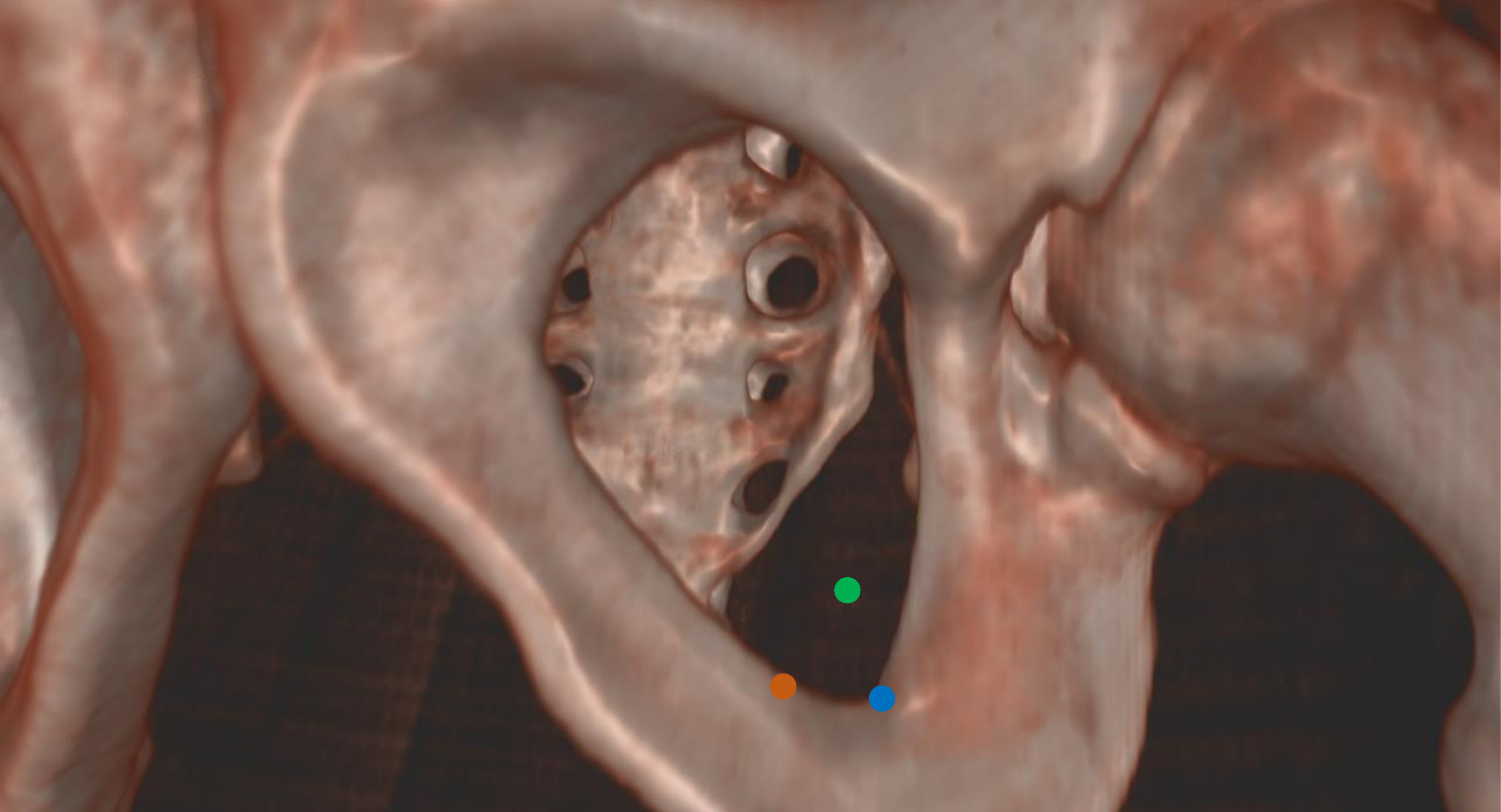}
\caption{
An original landmark (blue), the corresponding $p_\omega^{rays}$ (green), and the reprojected $m'_\omega$ (orange) for one landmark shown for one patient.
}\label{fig_refinement}
\end{figure}
The phase consists of three steps, as described below.\par
\color{black} 

\textbf{\textit{Phase}} \texttt{2.a} \textbf{\textit{--- Ray back-projection:}}
DRRs are generated from \ctpat using the same DRR generators, but a different smaller set of $K$ poses denoted as $S^2$.
The network $\mathbb{N}$ is then used to infer the position of the projections of the landmarks~$\mathcal{M}$. 
Due to the known geometry of the generated DRRs, each detected projection of a $m_\omega$ can be backprojected to a ray $r_\omega^k$, with $k=[1, 2, \dots, K]$ in 3D space. The ray spans between the landmark's projected location on the detector and the virtual X-ray source.\\
\indent\textbf{\textit{Phase}} \texttt{2.b} \textbf{\textit{--- Landmark refinement:}} The network~$\mathbb{N}$ has an inherent prediction inaccuracy, and hence the 3D rays~~$r_\omega^k$ resulting from the backprojection do not perfectly intersect at one point.
Therefore a new set of refined landmarks $\mathcal{M'} = [m'_1, m'_2, \dots, m'_\Omega]$ is generated, fulfilling the two following criteria: each new landmark $m'_\omega$ shall stay as close as possible to the barycenter of the intersection of all the rays~$r_\omega^k$, and be located on the bone surface.
The purpose of this refinement is to ensure that the final landmarks $m'_\omega$ describe meaningful anatomical regions, thus facilitating learning and registration. An example showing an original landmark, the corresponding barycenter of the intersection of rays and the refined position for one landmark on one patient is shown in Fig.~\ref{fig_refinement}.
This approach is independently conducted for all landmarks. For clarity purposes, it is subsequently described for a single landmark $m'_\omega$.\\
\indent First, the approximate barycenter~$p_\omega^{rays}$ of the intersection of all rays~$r_\omega^k$, is determined (Fig.~\ref{fig_refinement} green). 
Therefore, for each possible pair of given rays $\{r_{\omega}^{k_1}, r_{\omega}^{k_2}\}$, the closest equidistant point $p_\omega^{k_1, k_2}$ is calculated.
When the smallest distance between the two rays is larger than a constant threshold~$\tau$, the corresponding point~$p_\omega^{k_1, k_2}$ is discarded.
The coordinates of the barycenter $p_\omega^{rays}$ are then defined as the median $x$, $y$, and $z$ coordinates of all valid $p_\omega^{k_1, k_2}$ points.\\
\indent Then $p_\omega^{rays}$ is projected onto the bone surface, yielding the point~$m'_\omega$ (Fig.~\ref{fig_refinement} orange).
For this, the volume \ctpat is thresholded (Hounsfield units in $[200, 500]$ are mapped to \texttt{one} --- bone, all other values are mapped to \texttt{zero} --- background). 
Then, a contour detection algorithm is used to extract the bone surface from the thresholded volume~\cite{suzuki1985topological}.
Finally, the point $m'_\omega$ is determined using a sphere-growing scheme starting from $p_\omega^{rays}$ yielding the closest point on the bone surface.\\
\indent\textbf{\textit{Phase}} \texttt{2.c} \textbf{\textit{--- Patient-specific re-training:}}
The two landmark sets $\mathcal{M}$ and $\mathcal{M'}$ do not necessarily contain the same real-world points. Hence network re-training is necessary.
Therefore, a new network $\mathbb{N}'$ is automatically trained to detect the projections of the refined landmarks $\mathcal{M'}$. 
The weights of $\mathbb{N}'$ are initialized with the weights of $\mathbb{N}$.
To enable patient-specific re-training, the DRRs are only generated from \ctpat (as opposed to phase~\texttt{1}). The DRRs are generated from the same poses as during phase~\texttt{1} ($S^{1}$) and the same hyperparameters are used.\par
\textbf{\textit{Phase}} \texttt{2.d} \textbf{\textit{--- Pose-dependent Weights:}} The automation of all the previous steps does not only allow patient-specific re-training, but it also allows for an estimation of how accurate $\mathbb{N}'$ is at detecting the projections of the $\mathcal{M}'$ from different poses. This can be pre-computed in this phase, and then utilized during the initialization in order to obtain a better pose. Therefore, we introduce a set of pose-dependent weights computed using a new set of DRRs generated from \ctpat using a new set of poses denoted as $S^3$. After retraining, an inference is conducted on all the DRRs of this set, yielding the vector $\hat{\textbf{x}} = \{\hat{x}^{1}, \hat{x}^{2}, ... \hat{x}^{n}\}$, where each $\hat{x}^{i}$ is a set of detected locations for one pose. If a landmark $j$ is not detected, the corresponding values $\hat{x}^{i}_{j}$ are set to $(-1, -1)$. The corresponding ground truth vector, $\hat{\textbf{y}} = \{\hat{y}^{1}, \hat{y}^{2}, ... \hat{y}^{n}\}$, where each element corresponds to the projection of the set $\mathcal{M}'$ on one X-ray, is calculated using the known detector geometry. $\hat{\textbf{x}}$ and $\hat{\textbf{y}}$ are used to calculate distances for each pose using equation (\ref{eq:simil}) yielding the vector $\hat{\textbf{w}} = \{\hat{w}^{1}, \hat{w}^{2}, ... \hat{w}^{n}\}$. Each element is a vector of distances, that represent how accurate $\mathbb{N}'$ is at detecting each landmark at a specific DRR with a known pose. \\
\begin{equation} \label{eq:simil}
d(X, Y) = \begin{cases}
    40, & \text{if } X < 0 \lor Y < 0 \land X \neq Y\\
    \left\Vert X - Y\right\Vert, & \text{otherwise}
\end{cases}
\end{equation}
\indent The value $40$ was picked since it was the largest \revisionDel{distance} observed distance for the validation patient.\par
When analyzing Eq.~(\ref{eq:simil}) it is important to remember that $X$ or $Y$ are only negative if the corresponding landmark was not detected in the prediction or ground truth respectively.\par
\textbf{Phase \texttt{3} --- Intraoperative registration:}
This phase is conducted during the actual intervention.
For any new X-ray image (with unknown pose), it computes the transformation matrix~\t, which is used as an initialization for a given registration method~\r. First, the X-ray image is used as an input for the network $\mathbb{N}'$ to obtain the locations of the projections of the landmarks $\mathcal{M}'$ on the X-ray scan.\par
\textbf{\textit{Standard PnP:}} The standard way to compute the camera pose from a set of $2D-3D$ correspondences is to solve the corresponding Perspective-n-Point problem by iteratively minimizing the reprojection error $e_{re}$ between the $2D$ points and the corresponding $3D$ points using equation~(\ref{eq:PnP}). 
\begin{equation} \label{eq:PnP}
e_{re}(R, T, x, X) = \sqrt{\sum_{n=1}^{N}(x_n - x'_{n}(R,T,X))^2} 
\end{equation}
\indent Where $N$ is the number of detected landmarks, $R$ and $T$ are the current estimates for the rotation and translation of the C-arm, respectively, $x$ and $X$ are the vector of detected $2D$ points and $3D$ points, respectively and $x'(R, T, X)$ returns the projection of the $3D$ points according to the current estimate of $R$ and $T$. This is then minimized using a Levenberg-Marquardt optimizer, while the initial $R$ and $T$ are computed using a Direct Linear Transform algorithm.\par
\textbf{\textit{Modified PnP:}} In order to improve the accuracy of the PnP solver, we propose to incorporate pose-dependent weights. These weights are calculated using the vector of distances, $\hat{\textbf{w}}$, from phase 2.d. Intuitively, these distances encode how accurate $\mathbb{N}'$ is at predicting the projections of the $\mathcal{M}'$ for different poses, and is used to enhance the PnP algorithm as follows. First, we compute the nearest $k$ neighbours between the X-ray to register and all the poses of the DRRs created in 2.d\revisionDel{.} ($S^{3}$), using equation~(\ref{eq:simil}). This equation is used as a distance metric, taking as input the detected landmark locations for both the current X-ray and the DRRs. Second, the distance vectors $\hat{w}^{i}$ corresponding to the $k$ nearest neighbours are retrieved. These distances are then averaged, inverted and re-scaled to have mean one and standard deviation $0.5$ to convert them into weights. If the distance was zero, then the corresponding value is set to one before inverting. The inversion is done to ensure that small distances which correspond to better predictions, will be mapped to large weights in the cost function.
Finally, we propose a novel cost function for the PnP solver, which incorporates the pose-dependent weights, in order to place more emphasis on those landmarks which the network is good at detecting from this view and less emphasis on those it is bad at. On this function, higher emphasis will correspond to larger weights, since the cost function is minimized. Thereby, we replace equation~(\ref{eq:PnP}) with equation~(\ref{eq:weightedPnP}).
\begin{equation} \label{eq:weightedPnP}
e_{re}(R,T,\mathbb{W},x,X) = \sqrt{\sum_{n=1}^{N} \mathbb{W}_{n}(X)(x_n - x'_{n}(R,T,X))^2} 
\end{equation}
\indent where $\mathbb{W}_n(X)$ returns the pose-dependent weights for each detected point $X_n$. Each point $x_n$ is then weighted using the corresponding weight $\mathbb{W}_n$.

The computed initialization~\t is used to initialize a fine registration~\r, to accurately match the X-ray to \ctpat as detailed in Section~\ref{sec:experiments}.

\section{Experiments and Results}
\color{black} 

The structure of the experiments section is the following: First, an overview over the datasets used is given\revisionDel{.}\revisionAdd{;} Second, the experiment to determine the ideal hyperparameters during computation of the $p_\omega^{rays}$ is presented\revisionDel{.}\revisionAdd{;} Third, experiments to determine the right settings for $\mathbb{N}'$ and the unweighted initialization\revisionDel{.}\revisionAdd{;} Fourth, experiments regarding the pose-dependent weights are presented\revisionDel{.}\revisionAdd{;} Fifth, an experiment to evaluate the domain randomization is described\revisionDel{.}\revisionAdd{;} Last, experiments that evaluate the final registration are discussed.

\subsection{Datasets}

\begin{figure}[b]
\includegraphics[width=0.5\textwidth]{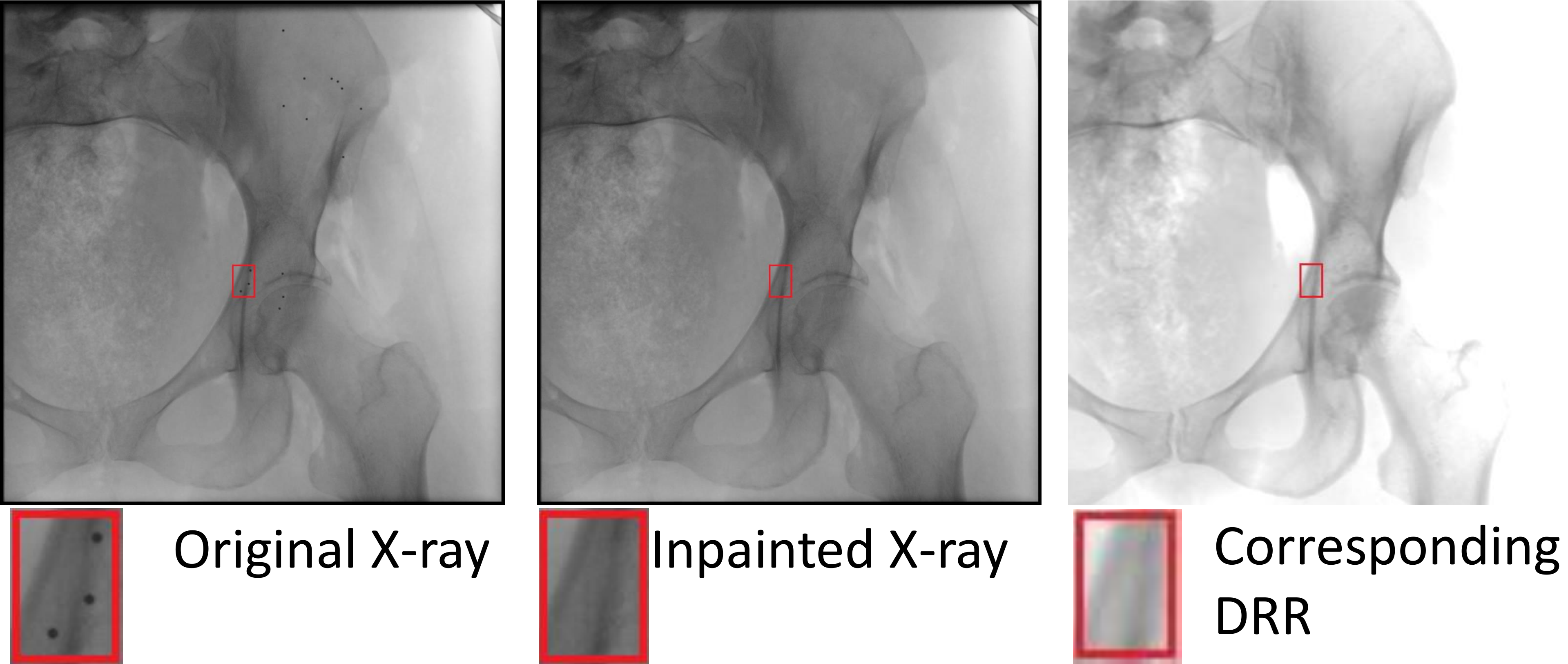}
\caption{
An original cadaver X-ray, its inpainted counterpart and a DRR generated from the corresponding view.
}
\label{fig_impainting}
\end{figure}

Three modalities are involved throughout the creation and testing of the method: CTs, X-rays, and DRR.\par
\textbf{\textit{CT Datasets:}} Four different datasets, which contain exactly one CT per patient, are used. For the development of the method: $\textbf{CT}_{train}$ (11 CTs) is used for training the network $\mathbb{N}$; $\textbf{CT}_{val}$ (1 CT) is used for validation during phase one and to determine the various hyperparameters used throughout the method. For testing and validation of the method: $\textbf{CT}_{synth}$ (six CTs) is used for assessing it's performance using synthetic data (DRRs); $\textbf{CT}_{cad}$ (two CTs, each from a cadaver) is used to test the method on real X-rays. The CTs for $\textbf{CT}_{train}$, $\textbf{CT}_{val}$, and $\textbf{CT}_{synth}$ are taken from the NIH Cancer Imaging Archive~\cite{roth2015new}.\par
\textbf{\textit{X-ray Dataset:}} \textcolor{black}{$Xray^{CT-cad}_{test}$ is the dataset with real X-rays taken from the two cadavers from $\textbf{CT}_{cad}$. It consists of $67$ X-rays from the first cadaver and $46$ from the second. In order to compute the ground truth registrations \revisionAdd{between} \revisionDel{for the} $\textbf{CT}_{cad}$ and \revisionDel{the} $Xray^{CT-cad}_{test}$, metal beads (BB) were injected into the two cadavers prior to acquisition of the X-rays. The locations of the BBs were then manually annotated and a ground truth registration was computed. Since the presence of the BBs results in an unrealistic experiment setup, the BBs were manually inpainted away from the CTs and all the X-rays. The difference can be seen in Fig.~\ref{fig_impainting}. The poses of all X-rays are denoted as  $S^{real}$.}\par
\textbf{\textit{\textcolor{black}{DRR Datasets:}}} \textcolor{black}{A DRR dataset is created by generating DRRs from a CT dataset using a set of poses, denoted as $S^{i}, i = 1 ... 4$. For every CT in the dataset, DRRs are generated using all the poses of a pose set. The movements of the artificial C-arm executed to create the sets $ S^{i}$ are shown in table~\ref{datasetsTable} a). Unlike the $S^{i}$, which are used to generate DRRs, $S^{real}$ corresponds to real X-rays from $\textbf{CT}_{cad}$ and hence their acquisition was not as structured. The DRR datasets and their corresponding combinations of CT sets and pose sets can be seen in table~\ref{datasetsTable} b).} 
$DRR^{CT-train}_{train}$ and $DRR^{CT-val}_{val}$ are used for training and validation of $\mathbb{N}$. $\textbf{CT}_{synth}$ and $\textbf{CT}_{cad}$ are used for validating the system, and represent a new patient to be processed by the method. Hence, the scans created from them are used for the steps in phase 2 and 3. $DRR^{CT-synth/cad}_{retrain}$ and $DRR^{CT-synth/cad}_{val}$ serve for retraining and validation for the patient specific network and also for generating $\mathcal{M}'$. The images from $DRR^{CT-synth/cad}_{cluster}$ are used for computing the pose-dependent weights, and finally $DRR^{CT-synth/cad}_{test}$ and $Xray^{CT-cad}_{test}$ represent intraoperative scans that need to be registered, where the former ones are DRRs and the latter one are real X-rays.
\begin{table}
  \centering
\includegraphics[width=0.5\textwidth]{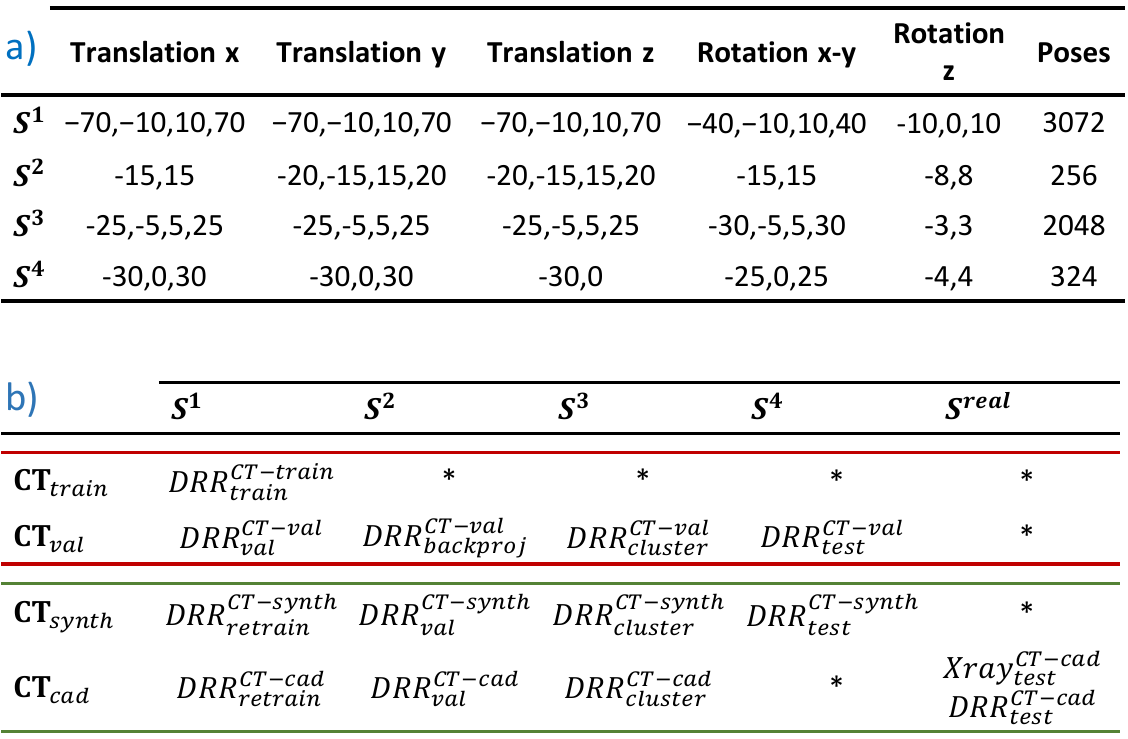}
\caption{Combinations of CT datasets and pose sets to generate the DRR datasets. The name ($DRR$ or $Xray$) indicates whether a set consists of real Xrays or DRRs, the superscript indicates the CTs from which the images were acquired and the subscript indicates the purpose of the set. Datasets generated from  $\textbf{CT}_{train}$ and  $\textbf{CT}_{val}$ (red) are used to develop phase 1. Datasets generated from $\textbf{CT}_{synth}$ and  $\textbf{CT}_{cad}$ (green) are used to validate the system as if a new patient was to be registered in a real settings.\\
}
\label{datasetsTable}
\end{table}
\color{black} 
\subsection{Imaging Parameters}

The X-rays were acquired using a Siemens Cios Fusion C-arm (Siemens Healthineers, Forchheim, Germany) with a flat panel detector. The system has a detector width of $384~mm$ by $384~mm$ and a source to detector distance of $1200~mm$.
\subsection{Evaluating ray backprojection}
This experiment is used to determine the right parameters for the heatmap threshold $\mu$ and the distance threshold $\tau$ during computation of the $p^{rays}_{\omega}$. It is conducted using $DRR^{CT-val}_{backproj}$ and one DRR Generator ($Gen_{1}$).\par
For this experiment, and the subsequent ones, $\mathbb{N}$ was trained for $4$ epochs until convergence was reached using the dataset $DRR^{CT-train}_{train}$ for training and $DRR^{CT-val}_{val}$ for validation.

Then, the $p_{\omega}$ are computed with different settings for $\tau$ and $\mu$. Finally, the euclidean distance between the $p_{\omega}$ and the $m_{\omega}$ is computed. The results can be seen in Table~\ref{tab_clusterTableVal}. Based on the results, the ideal values are $\tau = 15~mm$ and $\mu = 0.7$. Therefore, these settings are used throughout the remainder of the experiments.
\begin{table}
  \centering
  
  \includegraphics[width=0.5\textwidth]{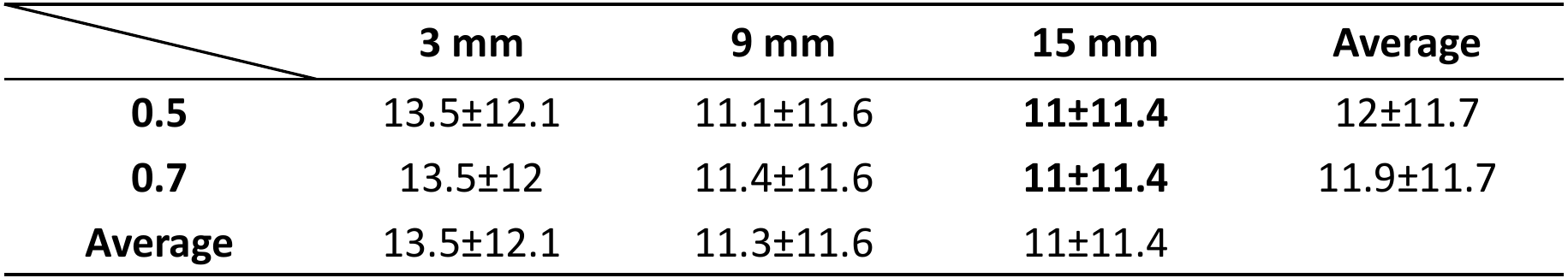}
\caption{
Average ($\pm$ standard deviation) euclidean distance (in mm) between the $m_{\omega}$ and $p^{rays}_{\omega}$ for $\textbf{D}^{CT}_{val}$ and different values of $\tau$ and $\mu$. 
}
\label{tab_clusterTableVal}
\end{table}
\subsection{Evaluating the unweighted initialization}
This subsection presents experiments to determine the right setting for training $\mathbb{N}'$, as well as to compute the initialization, ignoring the pose-dependent weights. Furthermore, this section shows the performance of the unweighted initialization. $\mathbb{N}'$ was trained using $DRR^{CT-test}_{retrain}$ and $DRR^{CT-val}_{retrain}$, respectively. The error metrics used throughout these experiments are the translation error in millimetres (mean euclidean distance between computed pose and ground truth pose) and the rotation error in degrees (mean absolute angle of the axis angle representation of $\mathbb{R}_{E} \cdot \mathbb{R}^{T}_{GT}$, where  $\mathbb{R}_{E}$ and $\mathbb{R}_{GT}$ are estimated and ground truth rotation).\par

First, several options for training $\mathbb{N}'$ are evaluated using $DRR^{CT-val}_{test}$. This was done using one DRR generator ($Gen_{1}$) and without pose-dependent weights. Several heatmap thresholds $\mu$ and several settings are compared. The settings include:
\begin{itemize}
    \item \textbf{From Scratch Cluster:} The network $\mathbb{N}'$ is initialized randomly and trained to detect the projections of the $p^{rays}_{\omega}$.
    \item \textbf{Retrain Cluster:} The network $\mathbb{N}'$ is initialized with the weights from $\mathbb{N}$ and trained to detect the projections of the $p^{rays}_{\omega}$.
    \item \textbf{Ground Truth Cluster:} Instead of using predictions from $\mathbb{N}'$, the initialization uses the $p^{rays}_{\omega}$ and their corresponding projections onto the detector.
    \item \textbf{From Scratch Bone:} The network $\mathbb{N}'$ is initialized randomly and trained to detect the projections of the $m'_{\omega}$.
    \item \textbf{Retrain Bone:} The network $\mathbb{N}'$ is initialized with the weights from $\mathbb{N}$ and trained to detect the projections of the $m'_{\omega}$.
    \item \textbf{Ground Truth Bone:} Instead of using predictions from $\mathbb{N}'$, the initialization uses the $m'_{\omega}$ and their corresponding projections onto the detector.
\end{itemize}

The results can be seen in Table~\ref{tab_fig_pnpTableVal}.
The \revisionDel{G} \revisionAdd{g}round \revisionDel{T} \revisionAdd{t}ruth settings are used to distinguish between errors from $\mathbb{N}'$ and errors from the subsequent PnP solver.\par
\begin{figure}
\includegraphics[width=0.5\textwidth]{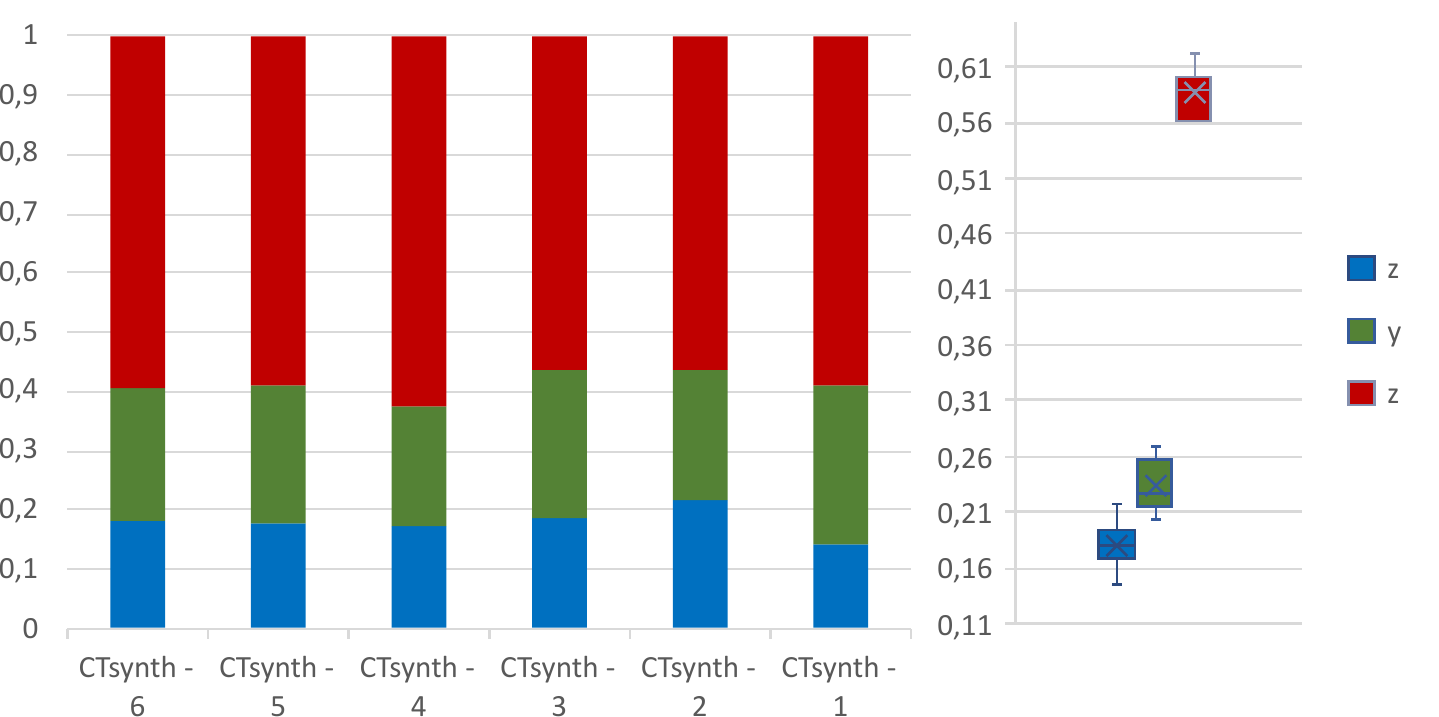}
\caption{
Decomposition of the translation error along the three camera axes for the Retrain Bone setting. The left picture shows the percentage of the error corresponding to each camera axis, separated by patient. The right picture shows a box plot with the error distribution along the full dataset. The error accumulated on the camera axis is bigger, as changes on this axis lead to little visual discrepancy.}
\label{fig_decompGraph}
\end{figure}

When comparing the results for the different $\mu$, a value of $0.8$ yields the best performance. The settings trained from scratch perform worse than the ones initialized with the weights of $\mathbb{N}$. When comparing the points $p^{rays}_{\omega}$ with the points $m'_{\omega}$ for the Retrain settings, the $m'_{\omega}$ outperform the $p^{rays}_{\omega}$. Hence the further experiments will be conducted using a $\mu$ of $0.8$ and the Retrain Bone setting.\\
\begin{table}
  \centering

    \includegraphics[width=0.5\textwidth]{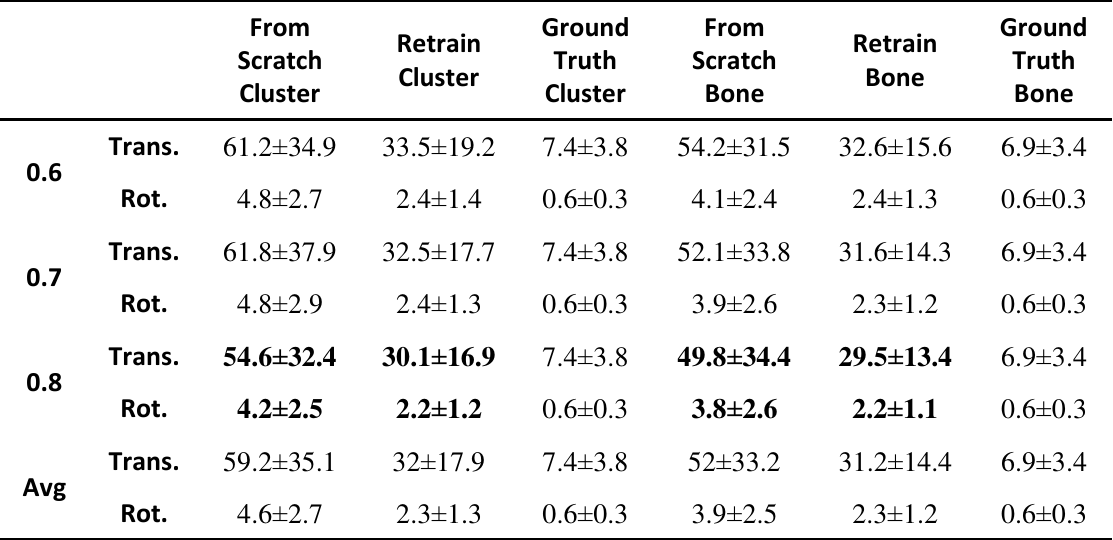}
    \caption{Average ($\pm$ standard deviation) camera pose error for the initialization without pose-dependent weights for different settings and different heatmap thresholds $\mu$. Translation error is in $mm$, and rotation error in degrees.}
    \label{tab_fig_pnpTableVal}
\end{table}
\indent Table~\ref{fig_pnpTable} shows the translation and rotation error for the initialization without pose-dependent weights for the dataset $DRR^{CT-synth}_{test}$, separated per patient. The heatmap threshold $\mu$ was set to $0.8$.
Again, one DRR generator is used ($Gen_{1}$). Besides the previously mentioned settings, this table introduces a new setting denoted as "No Retrain". This refers to a combination of the $p^{rays}_{\omega}$ and the predictions from $\mathbb{N}$. In other words, this refers to the method without $\mathbb{N}'$. As can be seen, this setting performs substantially worse than the others, justifying the need for $\mathbb{N}'$. When comparing Retrain Bone and Retrain Cluster, Retrain Bone performs better, thus confirming the previous experiment. Fig.~\ref{fig_decompGraph} shows the decomposition of the translation error along the three camera axes for the Retrain Bone setting. As expected, the error along the camera z-axis, which is the principal axis of the camera is larger than for the other two axes. This is because a motion along the principal axis leads to little visual change.


\begin{table}[]
    \includegraphics[width=0.5\textwidth]{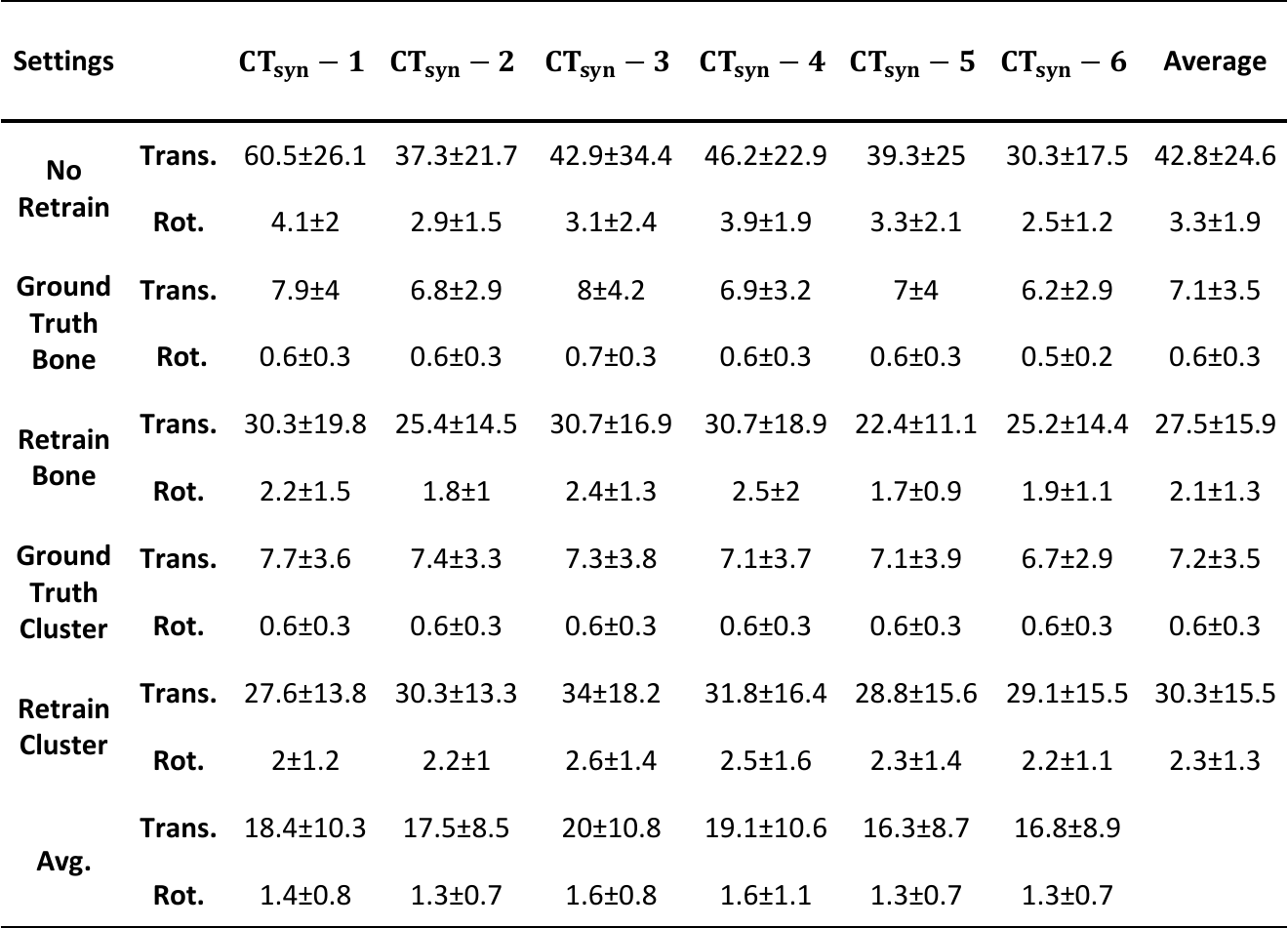}
    \caption{Average ($\pm$ standard deviation) camera pose error for the initialization without pose-dependent weights for different settings for each patient in $\textbf{CT}_{synth}$. The "No Retrain" setting is ignored when computing the average per patient. The heatmap threshold $\mu$ was set to $0.8$. Translation error is in $mm$, and rotation error in degrees.}
    \label{fig_pnpTable}
\end{table}


When comparing the numbers with the results from~\cite{esteban2019XrayReg}, which is the paper extended by this paper, it can be seen that the translation errors are larger, whereas the rotation errors are smaller. This is due to the Pelvis being placed further away from the radiation source in this work. The higher distance corresponds to more clinically meaningful poses. As expected, this leads to larger translation and smaller rotation error.\par
\begin{figure}
\includegraphics[width=0.49\textwidth]{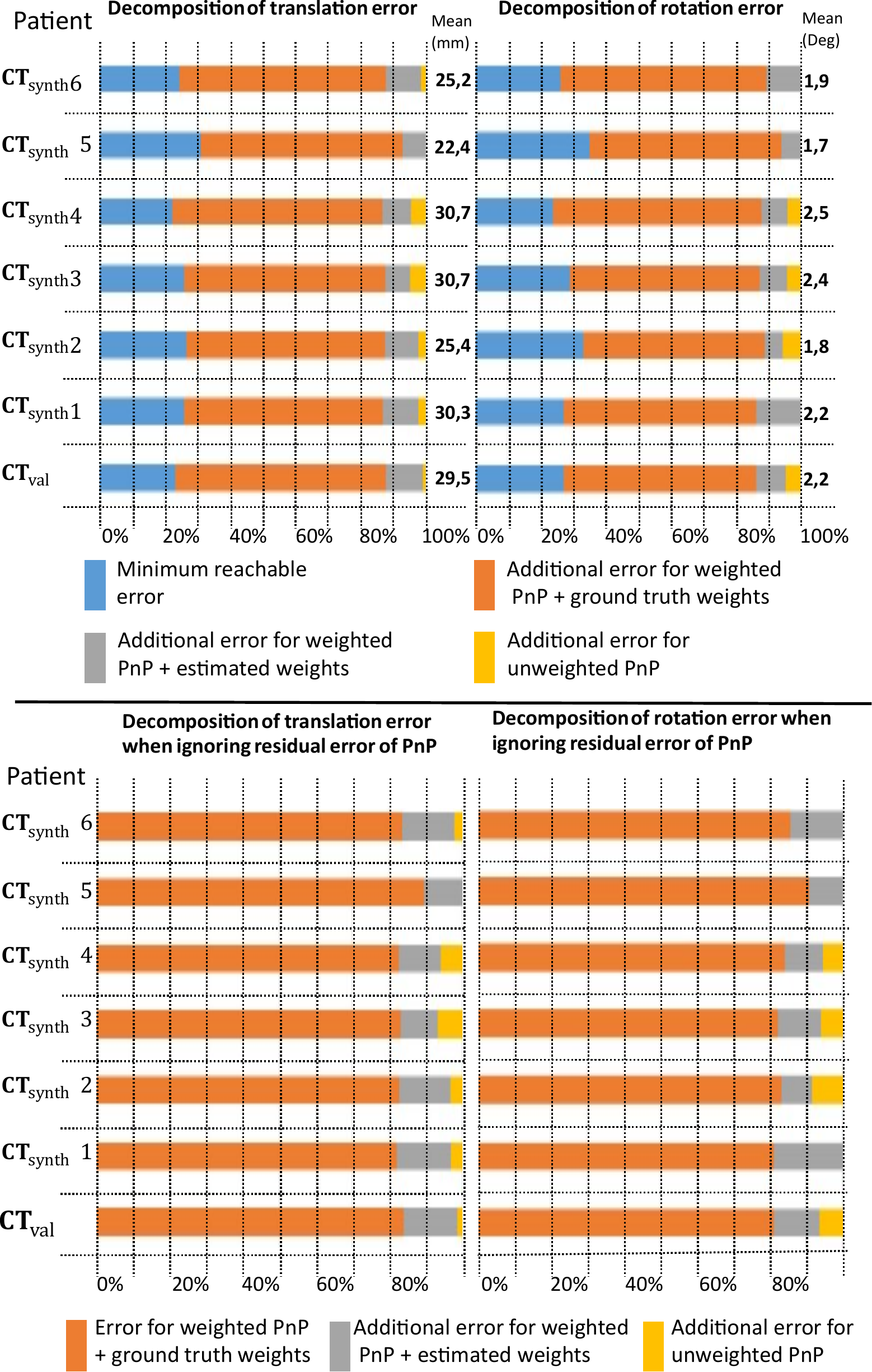}
\caption{
Error improvement for the translation and rotation error between computed pose and ground truth pose for the datasets $DRR^{CT-val}_{test}$ and $DRR^{CT-test}_{test}$, separated per patient.
The blue bar shows the error for an unweighted PnP using the ground truth correspondences between the $m'_{\omega}$ and their projections. Thereby, it shows the minimum possible error. The orange bar show the error for a weighted PnP where the weights are computed using the errors between the estimations from $\mathbb{N}'$ and the ground truth projections. The grey bar show the error for a weighted PnP using the proposed method to estimate the weights. The yellow bar shows the error for an unweighted PnP. The second row shows the decomposition of the error when ignoring the errors corresponding to the unweighted PnP with ground truth projections. The heatmap threshold $\mu$ was set to $0.8$.  
}
\label{fig_pnpSynthTable}
\end{figure}
When analyzing the errors of the \revisionDel{G} \revisionAdd{g}round \revisionDel{T} \revisionAdd{t}ruth settings, it is important to keep in mind the nature of the PnP solver, which uses a DLT initialization followed by a Levenberg-Marquardt based optimization of the parameters. Therefore, the algorithm does not necessarily find the optimal parameter set.\newpage

\subsection{Evaluating the pose-dependent weights}

\color{black}
This subsection describes the experiments carried out to compare the pose-dependent PnP to an unweighted PnP. It uses the same error metrics and training scheme for $\mathbb{N}'$ as the previous experiment. The pose-dependent weights are computed using \revisionDel{the} $DRR^{CT-test}_{cluster}$ and $DRR^{CT-val}_{cluster}$, respectively.

First, an experiment is conducted to test the theoretical boundaries of the pose-dependent weights. This demonstrate the maximum possible error improvement when using our method. Therefore, the notion of \revisionDel{G} \revisionAdd{g}round \revisionDel{T} \revisionAdd{t}ruth weights is introduced. These refer to weights that are computed for each sample using the prediction error of the landmarks for that specific sample. This serves as an upper bound on the performance of the algorithm, as during inference, the prediction errors are not known and therefore approximated using the errors of the k nearest neighbours. The result for this experiment for \revisionDel{the} $DRR^{CT-val}_{test}$ and $DRR^{CT-test}_{test}$ can be seen in Fig.~\ref{fig_pnpSynthTable} and the absolute values in Table~\ref{fig_weightedPnPTable-2}.

\color{black}

\begin{table*}[htbp]
    \includegraphics[width=1\textwidth]{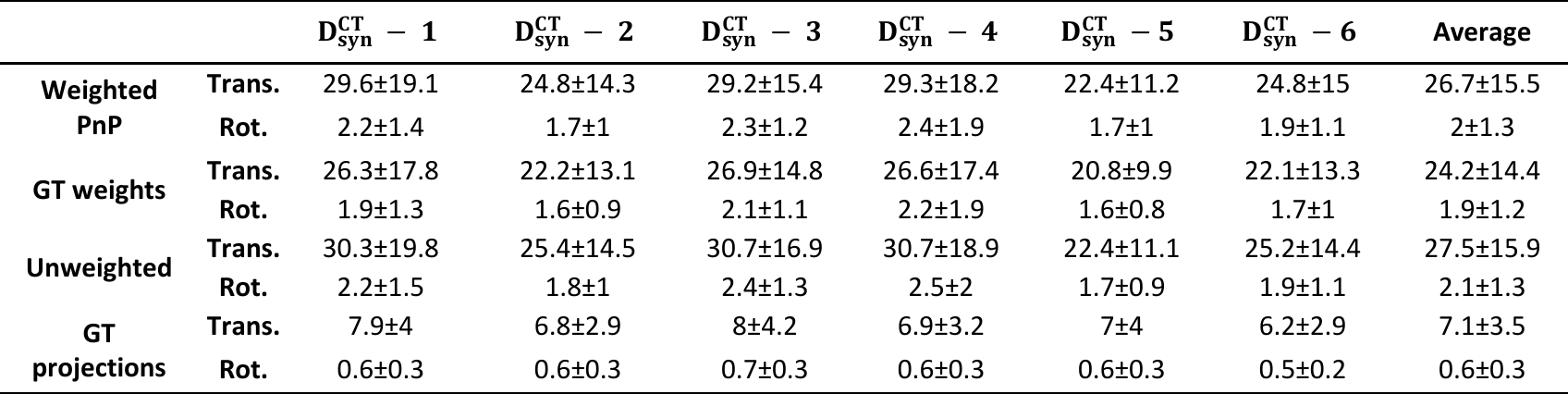}
    \caption{Average camera pose error for $DRR^{CT-synth}_{test}$ using weighted PnP with $11$ nearest neighbours, weighted PnP with ground truth weights, unweighted PnP and ground truth projections. Translation error is in $mm$, and rotation error in degrees.}
    \label{fig_weightedPnPTable-2}
\end{table*}


\begin{table}[t]
    \includegraphics[width=0.5\textwidth]{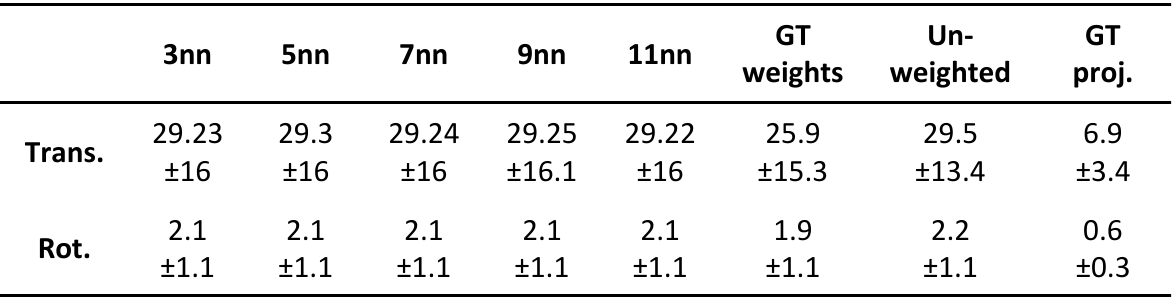}
    \caption{Average camera pose error for $DRR^{CT-val}_{test}$ for weighted PnP, where the weights are computed using a different number of neighbours, weighted PnP with ground truth weights, unweighted PnP, and unweighted PnP with ground truth projections. Translation error is in mm and rotation error in degrees. \revisionAdd{The nn stands for nearest neighbours.}}
    \label{fig_weightedPnPTable-1}
\end{table}

The blue bar in Fig.~\ref{fig_pnpSynthTable} shows the residual error of PnP. This error is the residual that results from running the PnP algorithm on a set of 3D points and their exact projections. This serves as an upper limit, as the weighing scheme can only counteract errors due to inaccuracies in the estimation of the landmarks. The grey bar shows the improvement when going from unweighted PnP to weighted PnP using \revisionDel{G} \revisionAdd{g}round \revisionDel{T} \revisionAdd{t}ruth weights. Hence the second row in Fig.~\ref{fig_pnpSynthTable} shows the decomposition when ignoring the residual error, thereby focusing on the part of the error that can actually be improved. When looking at the total error, the average improvement ($\pm$ standard deviation) when using ground truth weights over unweighted PnP is $11.8\pm2.1\%$ for translation and $10.9\pm2.5\%$ for rotation. When ignoring the residual error of PnP, it is $16\pm2.5\%$ for translation and $15.4\pm3.1\%$ for rotation.\par
After looking at the theoretical boundaries, experiments are conducted to determine the actual performance. Table~\ref{fig_weightedPnPTable-2} \textcolor{black}{shows the results for the pose-dependent weights. First, the ideal number of neighbours for computing the weights} is determined. This was done using $DRR^{CT-val}_{test}$. Table~\ref{fig_weightedPnPTable-1} shows the errors for multiple number of neighbours. As can be seen, the error differs only slightly. For the following experiments, $11$ was chosen as number of neighbours, as it performs slightly better than the other candidates.\par
Table~\ref{fig_weightedPnPTable-2} shows the results on the dataset $DRR^{CT-synth}_{test}$ using $11$ as number of neighbours for the weighted PnP. The weighted PnP always performs equal or outperforms the unweighted. Furthermore, it can be seen that the two patient with the highest improvement when going from unweighted to weighted PnP, also had the highest error for unweighted PnP ($\textbf{CT}_{synth}-3$ and $\textbf{CT}_{synth}-4$).
The performance of the ground truth weights is never reached. This is due to the distance between the poses $S^{3}$ used for computing the neighbours and the poses $S^{4}$. The closer they get, the more the weights returned by the nearest neighbour scheme resemble the ground truth weights. So two ways to further increase performance could be to either increase the size of $S^{3}$, which would result in a higher runtime, or induce prior knowledge about the poses to be encountered during the surgery if available.\par
\begin{table}
\includegraphics[width=0.5\textwidth]{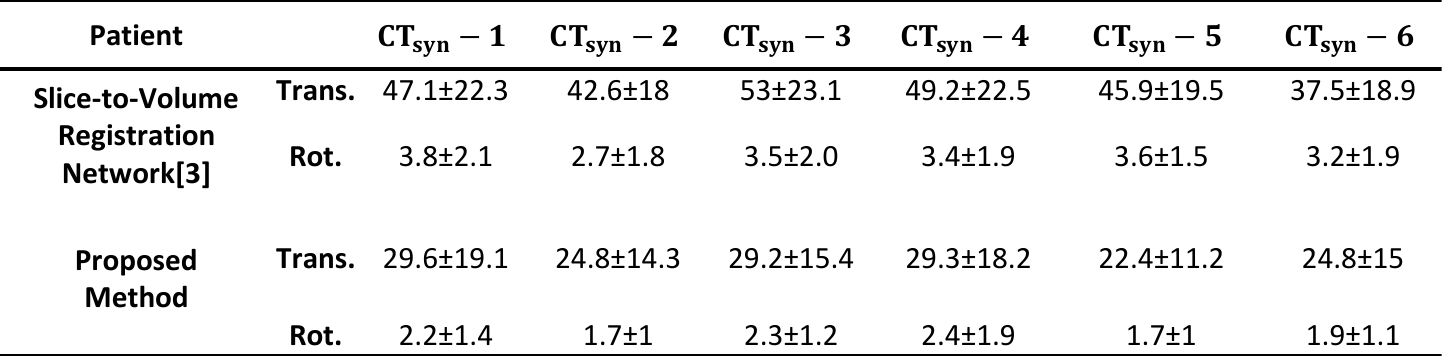}

\caption{
Comparison between the proposed initialization method and the previous state of the art fully automatic initialization method.
}
\label{fig_comparison}
\end{table}
Table~\ref{fig_comparison} shows the comparison of the proposed initialization method and the previous state of the art fully automatic initialization method~\cite{hou2017predicting}. Technically the method of \cite{hou2017predicting} can not be considered fully automatic, as it requires the annotation of three anatomical landmarks on CT. The first three landmarks of the set $\mathcal{M}$ were chosen. The method was trained using the dataset $DRR^{CT-train}_{train}$.
\subsection{Evaluating the Domain Randomization}

This subsection describes the experiments used to evaluate the domain randomization. Without the domain randomization, the method can only work on DRRs and not on real X-rays. The experiment is done using the $Xray^{CT-cad}_{test}$ and DRRs generated from the same poses, denoted as $DRR^{CT-cad}_{test}$. $\mathbb{N}'$ was trained using $DRR^{CT-cad}_{retrain}$.
\color{black} 

\begin{figure*}
\includegraphics[width=1.0\textwidth]{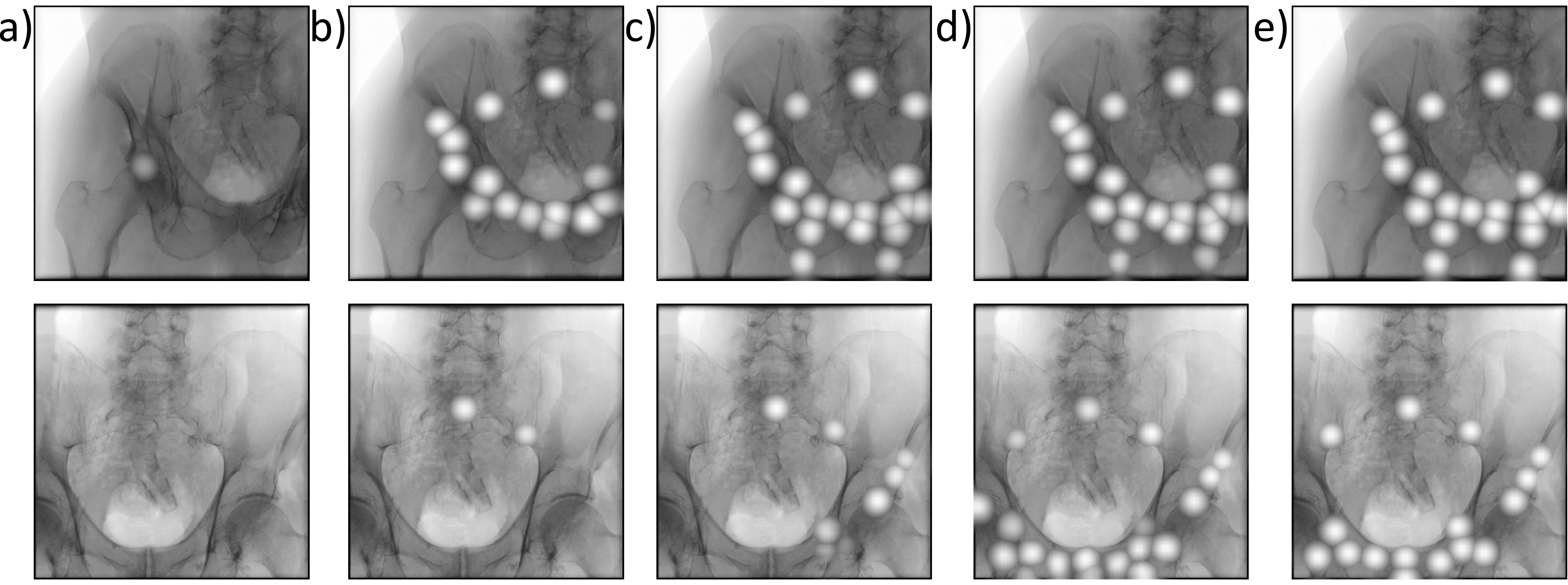}
\caption{
Examples of inferences on real X-rays for the different settings. a) Setting 2. b) Setting 3. c) Setting 4. d) Setting 5 e) Ground truth. Setting 1 is omitted, as no landmark was detected.
}
\label{fig_lotsOfXrays}
\end{figure*}


Five settings are evaluated. In setting 1 only $Gen_{1}$ is used. In setting 2, $Gen_1$ and $Gen_2$ are used. In setting 3, $Gen_1$, $Gen_2$ and $Gen_3$ are used.  For setting four, all four DRR generators are used. For setting five, all four DRR generators are used together with the post-processing scheme described in phase~\texttt{1}.\par
Example results can be seen in Fig.~\ref{fig_lotsOfXrays} and the quantitative numbers in Table~\ref{fig_realXrayTable}. As can be seen, the number of false positives is very low for all settings, both for DRRs and real X-rays. This is due to the high heatmap threshold $\mu$ of $0.8$. For correctly detected landmarks, the error for real X-rays is slightly higher than for DRRs for all settings. This is due to the network being trained using the same DRR generator that is used to generate the DRRs. Regarding the error on real X-rays, Settings 2, 3 and 4 slightly outperform setting 5. However, setting 5 has a much lower false negative rate ($20\%$ and $19.7\%$) than settings 4 ($41.4\%$ and $42.4\%$), 3 ($41.8\%$ and $75.1\%$) and 2 ($77.1\%$ and $98.7\%$). This can also be seen in Fig.~\ref{fig_lotsOfXrays}. When comparing setting 4 and 5, it can be seen that setting 5 performs better in terms of false negative rates for poses that are much closer to the anatomy than the poses in $S^{1}$ (Fig.~\ref{fig_lotsOfXrays} lower row). Except for setting 5, all settings have a significantly higher false negative rate for real X-rays than for DRRs. For settings 2, 3, 4 and 5, the error for real X-rays is only slightly worse than for DRRs. These two insights show that setting 5 enables the method to train purely on generated DRRs and then generalize to real X-rays. In general it can be seen that the more variation the network is exposed to during training, the better the generalization to real x-rays is.\par
\begin{table}
    \includegraphics[width=0.5\textwidth]{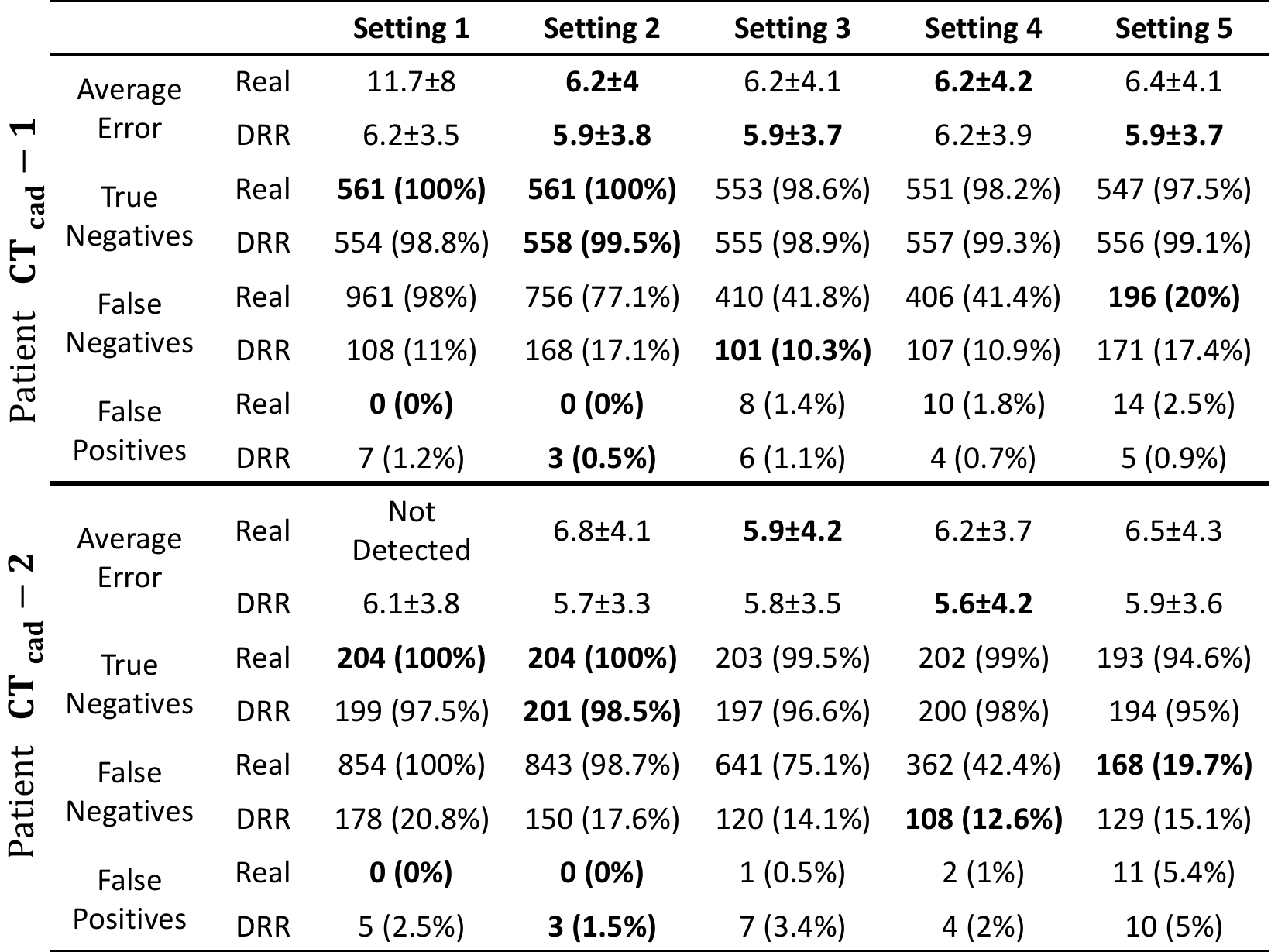}
    \caption{Quantitative results on the inference using real x-rays, using a heatmap threshold $\mu$ of $0.8$. Error denotes the average euclidean distance in millimetres ($\pm$ standard deviation). The numbers in brackets denote true negative rate, false negative rate and false positive rate, respectively. The detector has a size of $384$ by $384$ $mm$. Not Detected is used to denote that no landmark was detected.}
    \label{fig_realXrayTable}
\end{table}
When comparing settings 1-4 with the results from~\cite{unberath2019enabling} two things should be kept in mind. First, the false negative rate for settings 1-4 is very high. Second, $Gen_1$ and $Gen_2$ are used without the final log conversion, as stated in phase~$\texttt{1}$. Thereby, these are less realistic than the versions used by~\cite{unberath2019enabling}. Yet, this does not negatively affect the network's capabilities to run on real X-rays.
\subsection{Evaluating the final registration}
\label{sec:experiments}
This section describes the experiments carried out to show the final registration accuracy, thereby showing that the proposed initialization does indeed fall into the capture range of standard intensity-based algorithms.

\color{black} 

Therefore, the initialization method was used to initialize a standard intensity-based registration. The registration was done by pairing the BOBYQA~\cite{powell2009bobyqa} optimizer with the normalized cross correlation similarity metric. This registration is run three times. First, only the translation component is optimized as the rotation error of the initialization is comparably small. Second, only the rotation error is optimized. Finally, both translation and rotation are optimized jointly. The results can be seen in Fig.~\ref{fig_registration} and qualitative examples in Fig.~\ref{fig_chessboard}. A failure was defined as a translation error larger than $30~mm$. The target registration error (TRE) is computed using the $\mathcal{M}$ as they were not used during initialization. 
The scenario in the present paper is similar to the scenario of~\cite{grupp2019automatic}. When comparing the results, it can be seen that the present method has a higher error, but also a higher success rate. This is due to the more exclusive acceptance criterium chosen by~\cite{grupp2019automatic}.
\begin{table}
    \includegraphics[width=0.5\textwidth]{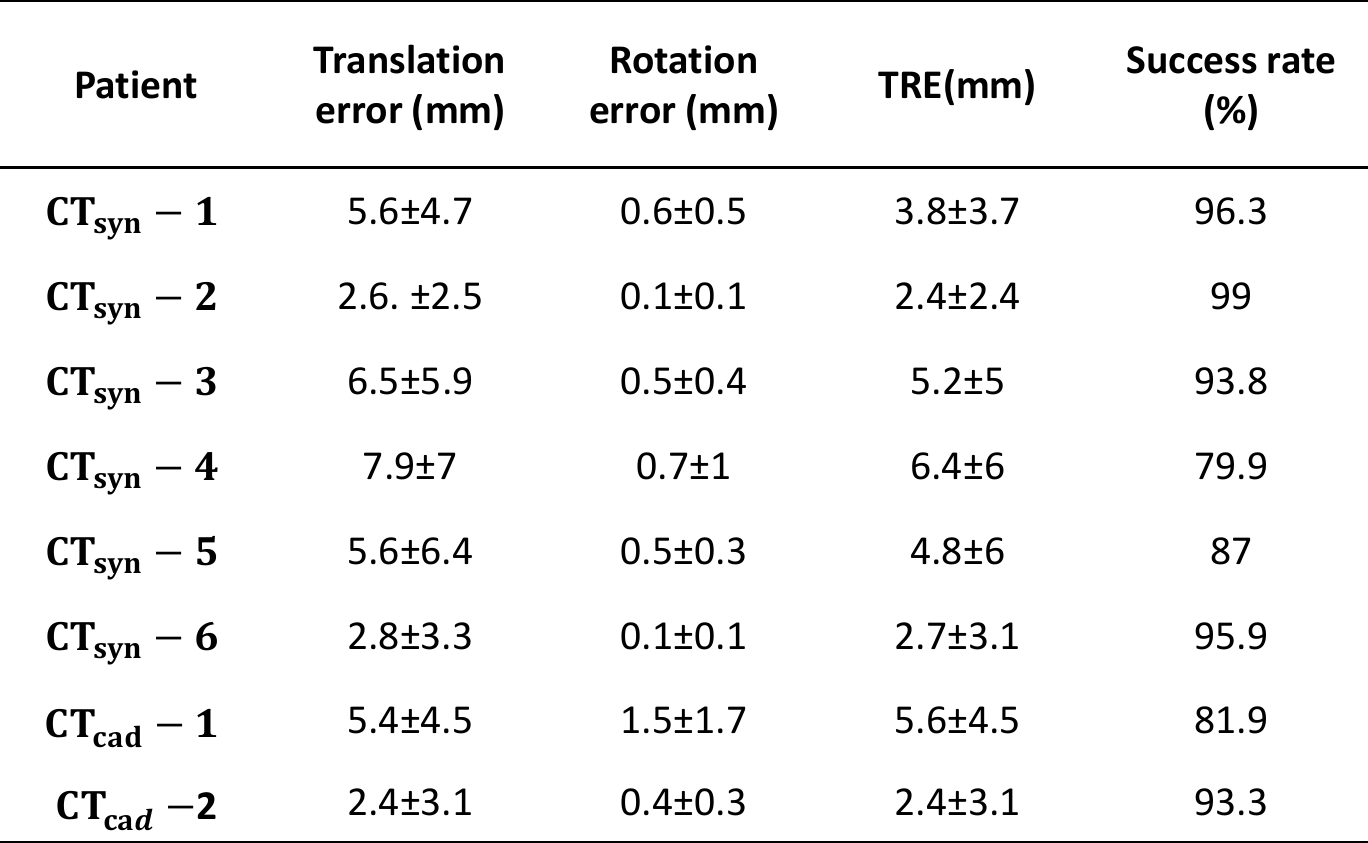}

\caption{
The error of the final registration method for $\textbf{DRR}^{CT-synth}_{test}$ and $\textbf{Xray}^{CT-cad}_{test}$. Translation error is in millimetres and rotation error in degrees. The target registration error is computed using the landmarks $\mathcal{M}$.
}
\label{fig_registration}
\end{table}
\section{Conclusion}
\label{sec:conclusion}
\color{black}
The present work introduced an initialization scheme for fully automatic X-ray to CT registration. The main novelty of the study is a patient-specific initialization step that can be used to initialize standard intensity-based registration methods.
Furthermore, a new augmentation scheme based on domain randomization was introduced to enable neural networks to be trained on simulated X-rays and better transfer to real X-rays. 

Future work could include extending the present work to other 2D to 3D modality combinations, such as X-ray to magnetic resonance imaging or Ultrasound to CT registration. 
Concluding, the proposed method paves the way for a new end-to-end fully-automatic X-ray to CT registration paradigm in the operating room.

\color{black}

\bibliographystyle{IEEEtran}
\color{black}
\bibliography{bib}

\begin{thebibliography}{10}
\providecommand{\url}[1]{#1}
\csname url@samestyle\endcsname
\providecommand{\newblock}{\relax}
\providecommand{\bibinfo}[2]{#2}
\providecommand{\BIBentrySTDinterwordspacing}{\spaceskip=0pt\relax}
\providecommand{\BIBentryALTinterwordstretchfactor}{4}
\providecommand{\BIBentryALTinterwordspacing}{\spaceskip=\fontdimen2\font plus
\BIBentryALTinterwordstretchfactor\fontdimen3\font minus
  \fontdimen4\font\relax}
\providecommand{\BIBforeignlanguage}[2]{{%
\expandafter\ifx\csname l@#1\endcsname\relax
\typeout{** WARNING: IEEEtran.bst: No hyphenation pattern has been}%
\typeout{** loaded for the language `#1'. Using the pattern for}%
\typeout{** the default language instead.}%
\else
\language=\csname l@#1\endcsname
\fi
#2}}
\providecommand{\BIBdecl}{\relax}
\BIBdecl

\bibitem{markelj2012review}
P.~Markelj, D.~Toma{\v{z}}evi{\v{c}}, B.~Likar, and F.~Pernu{\v{s}}, ``A review
  of {3D/2D} registration methods for image-guided interventions,'' \emph{Med.
  Im. Anal.}, vol.~16, no.~3, pp. 642--661, 2012.

\bibitem{van2010robust}
M.~J. Van~der Bom, L.~W. Bartels, M.~J. Gounis, R.~Homan, J.~Timmer, M.~A.
  Viergever, and J.~P.~W. Pluim, ``Robust initialization of 2{D}-3{D} image
  registration using the projection-slice theorem and phase correlation,''
  \emph{Med. Phys.}, vol.~37, no.~4, pp. 1884--1892, 2010.

\bibitem{hou2017predicting}
B.~Hou, A.~Alansary, S.~McDonagh, A.~Davidson, M.~Rutherford, J.~V. Hajnal,
  D.~Rueckert, B.~Glocker, and B.~Kainz, ``Predicting slice-to-volume
  transformation in presence of arbitrary subject motion,'' in \emph{Int. Conf.
  Med. Im. Comput. Comput. Assist. Interv.}, 2017, pp. 296--304.

\bibitem{miao2018dilated}
S.~Miao, S.~Piat, P.~Fischer, A.~Tuysuzoglu, P.~Mewes, T.~Mansi, and R.~Liao,
  ``Dilated fcn for {M}ulti-agent 2{D}/3{D} {M}edical {I}mage {R}egistration,''
  in \emph{Thirty-Second AAAI Conference on Artificial Intelligence}, 2018.

\bibitem{miao2016real}
S.~Miao, A.~Tuysuzoglu, Z.~J. Wang, and R.~Liao, ``Real-time 6dof pose recovery
  from x-ray images using library-based drr and hybrid optimization,''
  \emph{International Journal of Computer Assisted Radiology and Surgery},
  vol.~11, no.~6, pp. 1211--1220, 2016.

\bibitem{grupp2019automatic}
R.~B. Grupp, M.~Unberath, C.~Gao, R.~A. Hegeman, R.~J. Murphy, C.~P. Alexander,
  Y.~Otake, B.~A. McArthur, M.~Armand, and R.~H. Taylor, ``Automatic annotation
  of hip anatomy in fluoroscopy for robust and efficient 2d/3d registration,''
  \emph{International Journal of Computer Assisted Radiology and Surgery}, pp.
  1--11, 2020.

\bibitem{rackerseder2018initialize}
J.~Rackerseder, M.~Baust, R.~G{\"o}bl, N.~Navab, and C.~Hennersperger,
  ``Initialize globally before acting locally: {E}nabling landmark-free 3{D}
  {US} to {MRI} registration,'' in \emph{Int. Conf. Med. Im. Comput. Comput.
  Assist. Interv.}\hskip 1em plus 0.5em minus 0.4em\relax Springer, 2018, pp.
  827--835.

\bibitem{bier2018xray}
B.~Bier, M.~Unberath, J.-N. Zaech, J.~Fotouhi, M.~Armand, G.~Osgood, N.~Navab,
  and A.~Maier, ``X-ray-transform invariant anatomical landmark detection for
  pelvic trauma surgery,'' in \emph{Int. Conf. Med. Im. Comput. Comput. Assist.
  Interv.}\hskip 1em plus 0.5em minus 0.4em\relax Springer, 2018, pp. 55--63.

\bibitem{unberath2018deepdrr}
M.~Unberath, J.-N. Zaech, S.~C. Lee, B.~Bier, J.~Fotouhi, M.~Armand, and
  N.~Navab, ``Deepdrr--a catalyst for machine learning in fluoroscopy-guided
  procedures,'' in \emph{International Conference on Medical Image Computing
  and Computer-Assisted Intervention}.\hskip 1em plus 0.5em minus 0.4em\relax
  Springer, 2018, pp. 98--106.

\bibitem{unberath2019enabling}
M.~Unberath, J.-N. Zaech, C.~Gao, B.~Bier, F.~Goldmann, S.~C. Lee, J.~Fotouhi,
  R.~Taylor, M.~Armand, and N.~Navab, ``Enabling machine learning in
  x-ray-based procedures via realistic simulation of image formation,''
  \emph{International journal of computer assisted radiology and surgery},
  vol.~14, no.~9, pp. 1517--1528, 2019.

\bibitem{dou2019pnp}
Q.~Dou, C.~Ouyang, C.~Chen, H.~Chen, B.~Glocker, X.~Zhuang, and P.-A. Heng,
  ``Pnp-adanet: Plug-and-play adversarial domain adaptation network at unpaired
  cross-modality cardiac segmentation,'' \emph{IEEE Access}, vol.~7, pp.
  99\,065--99\,076, 2019.

\bibitem{zhang2020unsupervised}
Y.~Zhang, S.~Miao, T.~Mansi, and R.~Liao, ``Unsupervised x-ray image
  segmentation with task driven generative adversarial networks,''
  \emph{Medical Image Analysis}, vol.~62, p. 101664, 2020.

\bibitem{esteban2019XrayReg}
J.~Esteban, M.~Grimm, M.~Unberath, G.~Zahnd, and N.~Navab, ``Towards fully
  automatic x-ray to ct registration,'' in \emph{International Conference on
  Medical Image Computing and Computer-Assisted Intervention}.\hskip 1em plus
  0.5em minus 0.4em\relax Springer, 2019, pp. 631--639.

\bibitem{tobin2017domain}
J.~Tobin, R.~Fong, A.~Ray, J.~Schneider, W.~Zaremba, and P.~Abbeel, ``Domain
  randomization for transferring deep neural networks from simulation to the
  real world,'' in \emph{2017 IEEE/RSJ International Conference on Intelligent
  Robots and Systems (IROS)}.\hskip 1em plus 0.5em minus 0.4em\relax IEEE,
  2017, pp. 23--30.

\bibitem{wei2016convolutional}
S.-E. Wei, V.~Ramakrishna, T.~Kanade, and Y.~Sheikh, ``Convolutional pose
  machines,'' in \emph{IEEE Conf. Comput. Vis. Pattern Recognit.}, 2016, pp.
  4724--4732.

\bibitem{newell2016stacked}
A.~Newell, K.~Yang, and J.~Deng, ``Stacked hourglass networks for human pose
  estimation,'' in \emph{European conference on computer vision}.\hskip 1em
  plus 0.5em minus 0.4em\relax Springer, 2016, pp. 483--499.

\bibitem{suzuki1985topological}
S.~Suzuki and K.~Abe, ``Topological structural analysis of digitized binary
  images by border following,'' \emph{Comput. Vis. Graph. Image Process.},
  vol.~30, no.~1, pp. 32--46, 1985.

\bibitem{roth2015new}
H.~R. Roth, L.~Lu, A.~Seff, K.~M. Cherry, J.~Hoffman, S.~Wang, and R.~M.
  Summers, ``A new 2.5 {D} representation for lymph node detection in {CT},''
  \emph{The Cancer Imaging Archive}, 2015.

\bibitem{powell2009bobyqa}
M.~J. Powell, ``The bobyqa algorithm for bound constrained optimization without
  derivatives,'' \emph{Cambridge NA Report NA2009/06, University of Cambridge,
  Cambridge}, pp. 26--46, 2009.

\end{thebibliography}

\end{document}